\documentclass[11pt]{article}

\usepackage{times}
\usepackage{graphicx}
\usepackage{amssymb}
\usepackage{amsmath,amscd,amsfonts}
\usepackage{amsthm}
\usepackage[subnum]{cases}
\usepackage[utf8]{inputenc}
\usepackage[T1]{fontenc}
\usepackage[dvipsnames]{xcolor}
\usepackage{hhline}
\usepackage{multirow}
\usepackage{siunitx}
\usepackage{diagbox}
\usepackage{bbold}
\usepackage{authblk}
\RequirePackage[english,linesnumbered,algoruled]{algorithm2e}
\usepackage[a4paper, total={6.5in, 10in}]{geometry}

\def\ds{\displaystyle}
\def\R{\mathbb{R}}
\def\L{\mathbb{L}}
\def\E{\mathbb{E}}
\def\1{\textbf{1}}

\def\ds{\displaystyle}

\def\restriction#1#2{\mathchoice
              {\setbox1\hbox{${\displaystyle #1}_{\scriptstyle #2}$}
              \restrictionaux{#1}{#2}}
              {\setbox1\hbox{${\textstyle #1}_{\scriptstyle #2}$}
              \restrictionaux{#1}{#2}}
              {\setbox1\hbox{${\scriptstyle #1}_{\scriptscriptstyle #2}$}
              \restrictionaux{#1}{#2}}
              {\setbox1\hbox{${\scriptscriptstyle #1}_{\scriptscriptstyle #2}$}
              \restrictionaux{#1}{#2}}}
\def\restrictionaux#1#2{{#1\,\smash{\vrule height .8\ht1 depth .85\dp1}}_{\,#2}}

\renewcommand{\tilde}[1]{\widetilde{#1}}
\renewcommand{\leq}{\leqslant}

 \title{\bf Adaptive weighting of Bayesian physics informed\\ neural networks for multitask and multiscale\\ forward and inverse problems\thanks{Preprint submitted.}}

\author[1]{Sarah Perez}\author[2,3,4,5,6]{Suryanarayana Maddu} \author[2,3,4,5]{Ivo F.~Sbalzarini}\author[1]{Philippe Poncet}

\affil[1]{\small Universite de Pau et des Pays de l’Adour, E2S UPPA, CNRS, LMAP UMR CNRS-UPPA 5142, Pau, France.}
\affil[2]{Technische Universit\"{a}t Dresden, Faculty of Computer Science, N\"{o}thnitzer Str. 46, 01062 Dresden, Germany.}
\affil[3]{Max Planck Institute of Molecular Cell Biology and Genetics, Pfotenhauerstr. 108, 01307 Dresden, Germany.}
\affil[4]{Center for Systems Biology Dresden, Pfotenhauerstr. 108, 01307 Dresden, Germany.}
\affil[5]{Center for Scalable Data Analytics and Artificial Intelligence (ScaDS.AI), Dresden/Leipzig, Germany.}
\affil[6]{Center for Computational Biology, Flatiron Institute, 162 5th Avenue, New York NY 10010, USA.}

\begin{document}
\maketitle
 
\begin{abstract}
In this paper, we present a novel methodology for automatic adaptive weighting of Bayesian Physics-Informed Neural Networks (BPINNs), and we demonstrate that this makes it possible to robustly address multi-objective and multi-scale problems. BPINNs are a popular framework for data assimilation, combining the constraints of Uncertainty Quantification (UQ) and Partial Differential Equation (PDE). The relative weights of the BPINN target distribution terms are directly related to the inherent uncertainty in the respective learning tasks. Yet, they are usually manually set a-priori, that can lead to pathological behavior, stability concerns, and to conflicts between tasks which are obstacles that have deterred the use of BPINNs for inverse problems with multi-scale dynamics.

The present weighting strategy automatically tunes the weights by considering the multi-task nature of target posterior distribution. We show that this remedies the failure modes of BPINNs and provides efficient exploration of the optimal Pareto front. This leads to better convergence and stability of BPINN training while reducing sampling bias. The determined weights moreover carry information about task uncertainties, reflecting noise levels in the data and adequacy of the PDE model.
 
We demonstrate this in numerical experiments in Sobolev training, and compare them to analytically $\varepsilon$-optimal baseline, and in a multi-scale Lokta-Volterra inverse problem. We eventually apply this framework to an inpainting task and an inverse problem, involving latent field recovery for incompressible flow in complex geometries. \\

\end{abstract}

\section{Introduction}

Direct numerical simulation relies on appropriate mathematical models, derived from physical principles, to conceptualize real-world behavior and provide an understanding of complex phenomena. Experimental data are mainly used for parameter identification and a-posteriori model validation. However, a wide range of real-world applications are characterized by the absence of predictive physical models, notably in the life sciences. Data-driven inference of physical models has therefore emerged as a complementary approach in those applications \cite{maddu_stability_2019}.
The same is true for applications that rely on data assimilation and inverse modeling, for example in the geosciences. This has established data-driven models as complementary means to theory-driven models in scientific applications.

Depending on the amount of data available, several data-driven modeling strategies can be chosen. An overview of the state of the art in data-driven modeling, as well as of the remaining challenges, has recently been published~\cite{delia_machine_2022} with applications focusing on porous media research. It covers methods ranging from model inference using sparse regression \cite{brunton_sparse_id, fattahi_data-driven_2018, schaeffer_learning_2017, maddu_learning_2020}, where the symbolic structure of a Partial Differential Equation (PDE) model is inferred from the data, to equation-free forecasting models based on extrapolation of observed dynamics~\cite{Pathak_2018, Pathak_2020, maddu_stencil-net_2021}. 
Therefore, model inference methods are available for both physics-based and equation-free scenarios.

A popular framework combining both scenarios are Physics Informed Neural Networks (PINNs) \cite{RAISSI2019}. They integrate potentially sparse and noisy data with physical principles, such as conservation laws, expressed as mathematical equations. These equations regularize the neural network while the network weights $\theta \in \mathbb{R}^d$ and unknown equation coefficients $\Sigma\in\mathbb{R}^p$ are inferred from data. This has  enabled the use of PINNs as surrogate models, for example in fluid mechanics~\cite{oldenburg_geometry_2022, sun_surrogate_2020, poncetjfm2008}. Overall, PINNs provide an effective alternative to purely data-driven methods, since a lack of high-fidelity data can be compensated by physical regularization \cite{liu_multi-fidelity_2019, meng_multi-fidelity_2021}. 

Despite their effectiveness and versatility, PINNs can be difficult to use correctly, as they are prone to a range of training instabilities. This is because their training amounts to a weighted multi-objective optimization problem for the joint set of parameters $\Theta = \{\theta, \Sigma\}$,
\begin{equation}
    \label{multi_obj_loss}
    \ds \widehat{\Theta} = \mathrm{arg\,}\underset{\Theta}{\mathrm{min}}\, \sum_{k=0}^K \lambda_k \mathcal{L}_k(\Theta) ,
\end{equation}
where each term $\mathcal{L}_k(\Theta)$ of the loss function corresponds to a distinct inference task. For typical PINNs, these tasks include: data fitting, PDE residual minimization, boundary and initial condition matching, and additional physical constraints such as divergence-freeness of the learned field.
Proper training of this multi-task learning problem hinges on correctly setting the loss term weights $\lambda_k$ \cite{maddu_inverse_2022}. An unsuitable choice of weights can lead to biased optimization \cite{rahaman_spectral_2019}, vanishing task-specific gradients \cite{sener_multi-task_2018, kendall_multi-task_2018}, or catastrophic forgetting \cite{maddu_inverse_2022}. Automatically optimizing the loss weights, however, is not straightforward, especially in highly nonlinear and multi-scale problems.

The problem of how to tune the loss weights of a PINN is widely known and several potential solutions have been developed to balance the objectives \cite{chen_gradnorm_2018, Wang_2021, wang_when_2022, maddu_inverse_2022}. This offers criteria to impartially optimize the different tasks and provide a good exploration of the optimal Pareto front \cite{pareto_pinns}. While it improves reliability by reducing optimization bias, several open questions remain regarding the confidence in the predictions, noise estimates, and model adequacy \cite{NIPS2011_VI, meng_multi-fidelity_2021, yang_adversarial_2019}. These questions motivate a need for uncertainty quantification (UQ) to ensure trustworthy inference, extending PINNs to Bayesian inference in the form of BPINNs \cite{yang_b-pinns_2021, linka_bayesian_2022}. How to adapt successful PINN weighting strategies to BPINNs and integrate them with UQ, however, is an open problem.

BPINNs enable integration of UQ by providing posterior distribution estimates of the predictions — also known as Bayesian Model Averages \cite{wilson_bayesian_2020} — based on Markov Chain Monte Carlo (MCMC) sampling. One of the most popular MCMC schemes for BPINNs is Hamiltonian Monte Carlo (HMC), which provides a particularly efficient sampler for high-dimensional problems with smooth (physical) dynamics~\cite{betancourt_conceptual_2018}. Although HMC has been shown to be more efficient for BPINNs, its formulation implies potential energy that is related to the cost function of the PINN. The multi-objective loss of a PINN then directly translates to multi-potential energy for BPINN sampling. Therefore, it suffers from the same difficulties to avoid bias and provide an efficient exploration of the Pareto front.

This often causes HMC to not correctly explore the Pareto front during BPINN training. Efficient exploration of a high-dimensional Pareto front remains challenging for multi-task and multi-scale learning problems incorporating UQ and has not yet been addressed in the Bayesian case. The challenge arises because each term of the multi-potential energy is weighted within the Bayesian framework by parameters that relate to scaling, noise magnitude, and ultimately the inherent uncertainties in the different learning tasks \cite{kendall_multi-task_2018}. While these weights are recognized as critical parameters in the sampling procedure, they are mostly hand-tuned \cite{molnar_flow_2022, linka_bayesian_2022, meng_multi-fidelity_2021}, introducing hidden priors when the true uncertainties are not known. Appropriately setting these parameters is neither easy nor computationally efficient and can lead to either a biased estimation or a considerable waste of energy in ineffective tuning. Properly optimizing these weights is therefore essential to ensure that HMC samples from the posterior distribution around the Pareto front. This is not only required for robust BPINN training, but also for enhanced reliability of the UQ estimates. 

In order to robustly handle multi-task UQ inference in BPINNs, the open questions addressed in this article are: How can we automatically adjust the weights in BPINNs to efficiently explore the Pareto front and avoid bias in the UQ inference? How can we manage sensitivity to the noise distributions (homo- or hetero-scedastic) and their amplitude, without imposing hidden priors?

We start by characterizing potential BPINN failure modes, which are particularly prevalent for multi-scale or multi-task inverse problems. We then propose a modified HMC sampler, called Adaptively Weighted Hamiltonian Monte Carlo (AW-HMC), which avoids the problem by balancing gradient variances during sampling. We show that this leads to a weighted posterior distribution that is well suited to exploring the Pareto front. Our benchmarks show that this strategy reduces sampling bias and enhances the BPINNs robustness. In particular, our method improves the stability along the leapfrog steps during training, since it ensures optimal integration and frees the sampling from excessive time step decrease.  Moreover, it is able to automatically adjust the potential energy weights and with them, the uncertainties according to the sensitivity to noise of each term and their different scaling. This considerably improves the reliability of the UQ by reducing the need for hyperparameter tuning, including the prior distributions, and reducing the need for prior knowledge of noise levels or appropriate task scaling. We show that this improves BPINNs with respect to both the convergence rate and the computational cost. Moreover, we introduce a new metric for the quality of the prediction, quantifying the convergence rate during the marginalization step. We finally demonstrate that our proposed approach enables the use of BPINNs in multi-scale and multi-task Bayesian inference over complex real-world problems with sparse and noisy data.     

The remainder of this manuscript is organized as follows: In Sect. \ref{sec:BPINNs_overview_limits}, we review the general principles of BPINNs and the HMC sampler and characterize their failure mode in Sobolev training. Sect. \ref{sec:AW_HMC_for_BPINNs} describes the proposed adaptive weighting strategy for UQ using BPINNs. We validate this strategy in a benchmark with a known analytical solution in Sect. \ref{subsec:Sob_train_bench} and \ref{sec:2D_Sob_training}. We then demonstrate the effectiveness of the proposed AW-HMC algorithm on a Lokta-Volterra inverse problem in Sect. \ref{subsec:Lokta_Volterra}, focusing on a multi-scale inference of dynamical parameters. We then illustrate the use of AW-HMC in a real-world problem from fluid dynamics in Sect. \ref{sec:Results}. This particularly demonstrates successful inpainting of incompressible stenotic blood flow from  sparse and noisy data, highlighting UQ estimates consistent with the noise level and noise sensitivity. Finally, Sect. \ref{subsec:Inv_pb} considers an inverse flow problem in a complex geometry, where we infer both the flow regime (the inverse Reynolds number) and the latent pressure field from partial velocity measurements. We conclude and discuss our observations in Sect. \ref{sec:Conclusion}.

\section{From Uncertainty Quantification to Bayesian Physics-Informed Neural Networks: concepts and limitations}
\label{sec:BPINNs_overview_limits}

Real-world applications of data-driven or black-box surrogate models remain a challenging task. Predictions often need to combine prior physical knowledge, whose reliability can be questioned, with sparse and noisy data exhibiting measurement uncertainties. These real-world problems also suffer from non-linearity \cite{linka_bayesian_2022}, scaling \cite{cobb_scaling_2021}, and stiffness \cite{ji_stiff-pinn_2021} issues that can considerably impact the efficiency of the usual methodologies. This needs the development of data-driven modeling strategies that robustly address these issues.

At the same time, the need to build upon Bayesian inference raises the question in the research community of ensuring trustable intervals in the estimations. This is important for quantifying uncertainties on both the underlying physical model and the measurement data, although it may be challenging in the context of stiff, multi-scale, or multi-fidelity problems. Therefore, embedding UQ in the previous data-driven methodologies is essential to effectively manage real-world applications.

\subsection{HMC-BPINN concepts and principles}\label{subsec:BPINNs_review}

The growing popularity of Bayesian Physics-Informed Neural Networks \cite{yang_b-pinns_2021, linka_bayesian_2022, lin_multi-variance_2022, molnar_flow_2022, BPIExtreme} offers the opportunity to incorporate uncertainty quantification into PINNs standards, and benefit from their predictive power. It features an interesting Bayesian framework that claims to handle real-world sparse and noisy data and, as well, it bestows reliability on the models together with the predictions.

The basic idea behind a BPINN is to consider each unknown, namely the neural network and inverse parameters, $\Theta$, as random variables with specific distributions instead of single parameters as for a PINN. The different sampling strategies all aim to explore the posterior distribution of $\Theta$ 
\begin{equation}
    \label{post_dist_Bayes}
    P(\Theta |\mathcal{D}, \mathcal{M}) \propto  P(\mathcal{D}| \Theta) P(\mathcal{M}| \Theta ) P(\Theta) 
\end{equation} 
through a marginalization process, given some measurement data $\mathcal{D}$ and a presumed model $\mathcal{M}$, rather than looking for the best approximation satisfying the optimization problem \eqref{multi_obj_loss}. The posterior distribution expression \eqref{post_dist_Bayes} is obtained from Bayes theorem and basically involves a data-fitting likelihood term $P(\mathcal{D}| \Theta)$, a PDE-likelihood term $P(\mathcal{M}| \Theta )$ and a joint prior distribution $P(\Theta)$. These specific terms are detailed, case-by-case, in the applications, along with the different sections. The Bayesian marginalization then transfers the distribution of the parameters $\Theta$ into a posterior distribution of the predictions, also known as a Bayesian Model Average (BMA): 
\begin{equation}
    \label{BMA}
\underbrace{\mathcal{P}(y|x,\mathcal{D}, \mathcal{M})}_{\text{predictive BMA distribution}} = \int \underbrace{\mathcal{P}(y|x,\Theta)}_{\text{prediction for $\Theta$}} \underbrace{\mathcal{P}(\Theta|\mathcal{D}, \mathcal{M})}_{\text{posterior}} \mathrm{d}\Theta 
\end{equation}
where $x$ and $y$ respectively refer to the input (e.g spatial and temporal points) and output (e.g field prediction) of the neural network. In this equation, the different predictions arising from all the $\Theta$ parameters sampling \eqref{post_dist_Bayes} are weighted by their posterior probability and averaged to provide an intrinsic UQ of the BPINN output. Overall, BPINNs introduce a Bayesian marginalization of the parameters $\Theta$ which forms a predictive distribution \eqref{BMA} of the quantities of interest (QoI), namely the learned fields and inverse parameters. 

Different approaches were developed for Bayesian inference in deep neural networks including Variational Inference \cite{wu_deterministic_2022, liu_stein_2016} and Markov Chain Monte Carlo methods. A particular MCMC sampler based on Hamiltonian dynamics — the Hamiltonian Monte Carlo (HMC) — has drawn increasing attention due to its ability to handle high-dimensional problems by taking into account the geometric properties of the posterior distribution. Betancourt explained the efficiency of HMC through a conceptual comprehension of the method \cite{betancourt_conceptual_2018} and theoretically demonstrated the ergodicity and convergence of the chain \cite{betancourt_geometric_2017, HMC_Ergodicity_Betancourt}. From a numerical perspective, Yang et al. \cite{yang_b-pinns_2021} highlighted the out-performance of BPINNs-HMC formulation on forward and inverse problems compared to its Variational Inference declination. This has established HMC as a highly effective MCMC scheme for the BPINNs, both theoretically and numerically.

In the following, we briefly review the basic principles of the classical BPINNs-HMC and point out their limitations, especially in the case of multi-objective and multi-scale problems.   

The idea of HMC is to assume a fictive particle of successive positions and momenta $(\Theta, r)$ which follows the Hamiltonian dynamics on the frictionless negative log posterior (NLP) geometry. It requires the auxiliary variable $r$ to immerse the sampling of \eqref{post_dist_Bayes} into the exploration of a joint probability distribution $\pi(\Theta, r)$ in the phase space 
\begin{equation}
    \label{joint_dist}
    \pi(\Theta, r) \sim \mathrm{e}^{-H(\Theta, r)}.
\end{equation}
The latter relies on a particular decomposition of the Hamiltonian $H(\Theta, r) = U(\Theta) + K(r)$ where the potential and kinetic energies, $U(\Theta)$ and $K(r)$ respectively, are chosen such that 
\begin{equation}
    \label{joint_dist_marg}
    \pi(\Theta, r) \propto P(\Theta|\mathcal{D}, \mathcal{M}) \mathcal{N}(r|0, \mathbf{M}) 
\end{equation}
and the momentum follows a centered multivariate Gaussian distribution, with a covariance — or mass — matrix $\mathbf{M}$ often scaled identity. The Hamiltonian of the system is thus given by
\begin{equation}
    \label{Hamiltonian}
    H(\Theta, r) = U(\Theta)  + \frac{1}{2}r^T \mathbf{M}^{-1}r 
\end{equation}
where the potential energy directly relates to the target posterior distribution. This energy term is usually expressed as the negative log posterior $U(\Theta) = -\mathrm{ln} P(\Theta |\mathcal{D}, \mathcal{M})$, which results in a multi-potential as detailed in Sect. \ref{subsec:MO_paradigm}. This ensures that the marginal distribution of $\Theta$ provides immediate samples of the target posterior distribution
\begin{equation}
    \label{post_dist_potential}
    P(\Theta |\mathcal{D}, \mathcal{M})\sim \mathrm{e}^{-U(\theta)}
\end{equation} 
since an efficient exploration of the joint distribution $\pi(\Theta, r)$ directly projects to an efficient exploration of the target distribution, as described by Betancourt \cite{betancourt_conceptual_2018}. The HMC sampling process alternates between deterministic steps, where we solve for the path of a frictionless particle given the Hamiltonian dynamical system 
\begin{equation}
    \label{Ham_dyn_sys}
    \left\{\begin{array}{l}
    \mathrm{d}\Theta = \mathbf{M}^{-1}r\,\mathrm{d}t \\
    \mathrm{d}r = -\nabla U(\Theta) \mathrm\,{d}t, 
    \end{array}\right.
\end{equation}
and stochastic steps, where the momentum is sampled according to the previously introduced Gaussian distribution. As Hamilton's equation \eqref{Ham_dyn_sys} theoretically preserves the total energy of the system, each deterministic step is then constrained to a specific energy level while the stochastic steps enable us to diffuse across the energy level set for efficient exploration in the phase space. This theoretical conservation of the energy level set during the deterministic steps requires numerical schemes that ensure energy conservation. 

A symplectic integrator is thus commonly used to numerically solve for the Hamiltonian dynamics \eqref{Ham_dyn_sys}: the Störmer-Verlet also known as the leapfrog method.  However, these integrators are not completely free of discretization errors that may disrupt, in practice, the Hamiltonian conservation through the deterministic iterations. Hence, a correction step is finally added in the process to reduce the bias induced by these discretization errors in the numerical integration: this results in a Metropolis-Hasting criterion based on the Hamiltonian transition. This acceptance criterion tends to preserve energy by rejecting samples that lead to divergent probability transition. The exploration of the deterministic trajectories though remains sensitive to two specific hyperparameters managing the integration time: the step size $\delta t$ and the number of iterations $L$ used in the leapfrog method. Tuning these parameters can be challenging, especially if the posterior distribution presents pathological or high curvature regions \cite{betancourt_conceptual_2018}, yielding instability, under-performance, and poor validity of the MCMC estimators. Despite the use of numerical schemes that preserve the Hamiltonian properties, a conventional HMC-BPINN can be confronted with pathological discrepancies. 

To counteract these divergence effects, efforts have been put into developing strategies to either adaptively set the trajectory length $L$ \cite{hoffman_adaptive-mcmc_2021} while preserving detailed balance condition or use standard adaptive-MCMC approaches to adjust the step size $\delta t$ on the fly \cite{Adapt_step_size_HMC}. In this regard, one of the most popular adaptive strategies is the No-U-Turn sampler (NUTS) from Hoffmann and Gelman \cite{homan_no-u-turn_2014}. Nonetheless, these divergent trajectories indicate significant bias in the MCMC estimation even if such adaptive methods may offer an alternative to overcome them. This raises the question of the validity of this adaptation when facing multi-potential energy terms that lead to significantly different geometrical behaviors or different scaling in the posterior distribution. In fact, the adaptive strategy mostly tunes the leapfrog parameters so that the most sensitive term respects  the energy conservation, which may result in poorly-chosen hyper-parameters for the other potential energy terms, and then the whole posterior distribution. This reflects the limitations of such adaptive strategies that rely on adjusting the leapfrog hyperparameters. 

When these divergent pathologies become prevalent, another approach suggested by Betancourt \cite{betancourt_conceptual_2018} is to regularize the target distribution, which can become strenuous in real-world applications and lead to additional tuning. Nevertheless, it offers a great opportunity to investigate the impact of each learning task on the overall behavior of the target distribution and paves the way for novel adaptive weighting strategies. 

In the next sections, we focus particularly on the challenges arising from real-world multi-tasks and multi-scale paradigms. We show that present BPINN methods result in major failures in these cases and we identify the main pathologies using powerful diagnostics based on these divergent probability transitions.

\subsection{The multi-objective problem paradigm}
\label{subsec:MO_paradigm}

As for the issue of the multi-objective optimization problem in a PINN, sampling of the target posterior distribution \eqref{post_dist_Bayes} arising from a direct or inverse problem requires the use of a multi-potential energy term $U(\Theta)$. Furthermore, in real-world applications, we have to deal with sparse and noisy measurements whose fidelity can also cover different scales: this is the case of multi-fidelity problems with multi-source data \cite{liu_multi-fidelity_2019, meng_multi-fidelity_2021}.    

For sake of generality, we introduce a spatio-temporal domain $\Omega = \tilde\Omega\times \mathcal{T}$ with $\tilde\Omega\subset \R^n$, $n = 1,2,3$ and we assume a PDE system in the following form: 
\begin{equation}
    \label{PDE_sys}
    \left\{\begin{array}{ll}
    \mathcal{F}(u(t,x),\ \Sigma) &= 0, \qquad (t,x)\in \Omega\\
    \mathcal{H}(u(t,x),\  \Sigma) &= 0, \qquad (t,x)\in \Omega\\
    \mathcal{B}(u(t,x),\  \Sigma) &= 0, \qquad (t,x)\in \Omega^{\partial} := \partial\tilde\Omega\times\mathcal{T}\\
    \mathcal{I}(u(t,x),\  \Sigma) &= 0, \qquad (t,x)\in \Omega^{I} := \tilde\Omega\times\mathcal{T}_0
    \end{array}\right.   
\end{equation}
where $u$ is the principal unknown, $\mathcal{F}$ the main differential equation (e.g the Navier-Stokes equation), $\mathcal{H}$ an additional constraint (e.g incompressibility condition), $\mathcal{B}$ and $\mathcal{I}$ the boundary and initial conditions respectively, and $\Sigma$ the PDE model parameters, either known or inferred. Some partial measurements of the solution field $u$ may also be available in a subset $\Omega^u\subset\Omega$. Such a continuous description of the spatio-temporal domain is then discretized to enable the selection of the training dataset, which is used in BPINNs sampling.

We first define the dataset $\mathcal{D}$ of training data which is decomposed into $\mathcal{D} = \mathcal{D}^\Omega \cup \mathcal{D}^\partial \cup \mathcal{D}^I \cup \mathcal{D}^u$ and includes scattered and noisy measurements sampled in their respective sets $\Omega$, $\Omega^\partial$, $\Omega^I$, and $\Omega^u$. Regarding data corruption, we consider independent Gaussian noise for the sparse observations on $u$, such that $\mathcal{D}^u$ is defined as
\begin{equation}
\label{D_u}
    \mathcal{D}^u = \left\{  (t_i,x_i, u_i), \quad \text{s.t}\quad (t_i, x_i) \in\Omega^u  \quad \text{and} \quad u_i := u(t_i,x_i) + \xi_u(t_i,x_i), \, i=1...N^u \right\}
\end{equation}
where the noise $\xi_u \sim \mathcal{N}(0,\sigma_u^2I)$ and the standard deviation $\sigma_u$ might be estimated from the sensor fidelity, if accessible. The neural network component of the BPINN then provides a surrogate model of $u$ denoted $u_\Theta$ for each sample of the parameters $\Theta =  \{\theta, \Sigma\}$, whose prior distribution is referred to as $P(\Theta)$. The latter takes into account both the priors on the neural network parameters $\theta$, which are assumed to be centered and independent Gaussian distributions, and the priors on the model parameters $\Sigma$, so that $P(\Theta) = P(\theta)P(\Sigma)$ under the independence condition. In the case of a forward problem, where the PDE model parameters are prescribed, the prior distribution reduces to $P(\theta)$. When some measurements of the unknown are available, meaning $\mathcal{D}^u$ is not an empty set, which is the case in inverse or inpainting problems, then the surrogate model $u_\Theta$ should satisfy a data-fitting likelihood term in the Bayesian framework. This consists in quantifying, over the set $\mathcal{D}^u$, the fit between the neural network prediction and the training data defined by: 
\begin{equation}
    \label{Data_likelihood}
    P(\mathcal{D}^u | \Theta) \propto \prod_{i=1}^{N^u} \mathrm{exp}\left(\frac{-\left(u_\Theta(t_i,x_i) - u_i\right)^2}{2\sigma_u^2} \right).
\end{equation}
Similarly, the boundary conditions of the model output are imposed on the set $\mathcal{D}^\partial$
\begin{equation}
\label{D_bc}
    \mathcal{D}^\partial = \left\{  (t_i,x_i, \mathcal{B}(u_i)), \quad \text{s.t}\quad (t_i,x_i)\in\Omega^\partial \quad \text{and} \quad \mathcal{B}(u_i):= \mathcal{B}(u(t_i,x_i)) + \xi_b(t_i,x_i), \, i=1...N^\partial \right\}
\end{equation}
by satisfying the following boundary-likelihood term
\begin{equation}
    \label{BC_likelihood}
    P(\mathcal{D}^\partial | \Theta) \propto \prod_{i=1}^{N^\partial} \mathrm{exp}\left(\frac{-\left(\mathcal{B}(u_\Theta(t_i,x_i)) - \mathcal{B}(u_i)\right)^2}{2\sigma_b^2} \right).
\end{equation}
The noise sensitivity on the boundary condition term is also characterized by independent Gaussian distributions in the sense that $\xi_b \sim \mathcal{N}(0,\sigma_b^2I)$ where the standard deviation $\sigma_b$ needs to be estimated. Such a distinction between $\xi_u$ and $\xi_b$ is prescribed since there is no guarantee that the data corruption is uniform: in fact, the measurement distribution variances can differ locally when facing heteroscedastic noise. This is the case in geosciences, where data-driven modeling based on X-Ray microtomography images require special attention on this boundary noise estimation $\xi_b$. This is mainly due to the artifact limitations (e.g partial volume effect, edge-enhancement) that tend to enhance the blurring effects at the material interface and therefore impact the quantification of the medium effective properties, such as the permeability and micro-porosity \cite{perez_deviation_2022, alqahtani_super-resolved_2022}. The same holds for the initial condition with potentially a different sensitivity $\xi_i$. In a BPINN, the previous data-fitting terms are complemented with physical principles that regularize the neural network predictions, given the PDE system \eqref{PDE_sys}.

Concerning the PDE-likelihood term, the $\mathcal{D}^\Omega$ dataset is defined as the training points on which we force the PDE and the additional physical constraint to be satisfied by the surrogate modeling:
\begin{equation}
\label{D_pde}
    \mathcal{D}^\Omega = \left\{  (t_i,x_i)\in\Omega, \quad  \mathcal{F}(u_\Theta(t_i,x_i)) = \xi_f(t_i, x_i) \quad \text{and} \quad \mathcal{H}(u_\Theta(t_i,x_i)) = \xi_h(t_i, x_i), \, i=1...N^\Omega \right\}
\end{equation}
with $\xi_f$ and $\xi_h$ standing for the model uncertainty in both equations, which are usually unknown and can easily lead to physical model misspecification. According to these notations, a forward problem consists in $\mathcal{D}^u=\varnothing$ and $\Sigma$ is known to perform a direct prediction of the field $u_\Theta$ on $\Omega$ based only on the PDE physical assumptions. On the contrary, an inverse problem aims to infer $\Sigma$ using together the PDE model with the partial and noisy information $\mathcal{D}^u$ of the predictive field $u$. Finally, an inpainting problem relies on these partial measurements to complement and recover some missing information on the predictive field, in addition to the PDE-based priors.

Finally, the target posterior distribution of $\Theta$ \eqref{post_dist_Bayes} is decomposed according to the Bayes rule, into a sequence of multi-task likelihood terms —- involving data-fitting and PDE likelihood — and the priors:
\begin{equation}
    \label{post_dist_multi}
   P(\Theta |\mathcal{D}, \mathcal{M}) \propto  P(\mathcal{D}^u| \Theta)P(\mathcal{D}^\partial| \Theta)P(\mathcal{D}^I| \Theta)P(\mathcal{D}^\Omega, \mathcal{F}| \Theta)P(\mathcal{D}^\Omega, \mathcal{H}| \Theta) P(\theta)P(\Sigma) 
\end{equation} 
which results, for the HMC sampler, in the multi-potential energy  
\begin{equation}
    \label{multi_potential}
    \begin{aligned}
    U(\Theta) = &\frac{\left\|u_\Theta - u\right\|_{\mathcal{D}^u}^2}{2\sigma_u^2}
    + \frac{\left\|\mathcal{B}(u_\Theta) - \mathcal{B}(u)\right\|_{\mathcal{D}^\partial}^2}{2\sigma_b^2}
    + \frac{\left\|\mathcal{I}(u_\Theta) - \mathcal{I}(u)\right\|_{\mathcal{D}^I}^2}{2\sigma_i^2}\\
    & \qquad + \frac{\left\|\mathcal{F}(u_\Theta)\right\|_{\mathcal{D}^\Omega}^2}{2\sigma_f^2}
    + \frac{\left\|\mathcal{H}(u_\Theta)\right\|_{\mathcal{D}^\Omega}^2}{2\sigma_h^2}
    + \frac{\left\| \theta\right\|_{\R^d}^2}{2\sigma_\theta^2} 
    + \frac{\left\| \Sigma - \mu_\Sigma\right\|_{\R^p}^2}{2\sigma_\Sigma^2} 
    \end{aligned}
\end{equation}
according to equation \eqref{post_dist_potential}. The notation $\|\cdot\|$ refers to either the RMS (root mean square) norm — inherited from the functional $\L^2$-norm on the open set $\Omega$ — for the log-likelihood terms or to the usual Euclidean norm for the log-prior terms. In addition, the multi-potential \eqref{multi_potential} is written here, in a general framework, based on the prior assumptions $P(\theta)\sim\mathcal{N}(0,\sigma_\theta^2 I_d)$ and $P(\Sigma)\sim\mathcal{N}(\mu_\Sigma, \sigma_\Sigma^2 I_p)$. We note that the log-prior term can be regarded as a $\L^2$-regularization in the equivalent constrained optimization problem. Nonetheless, suitable selection of these prior distributions — hence appropriate tuning of the parameters $\sigma_\theta$, $\mu_\Sigma$, and $\sigma_\Sigma$ — is usually not straightforward and is time-consuming. Overall, equation \eqref{multi_potential} highlights that, even in a simple problem setup, a BPINN may face a potential energy term that closely resembles a weighted multi-objective loss appearing in a PINN, whose weights are mainly hand-tuned. 

Therefore, the main challenge is to sample near the Pareto-optimal solution such that the BPINNs provide efficient and reliable prediction and UQ. Otherwise, the risk is that the samples obtained gravitate around a local minimum, corresponding to one of the multi-potential terms at the cost of the others.

Secondly, while the standard deviations $\sigma_\bullet$ are critical parameters to select and are related to the uncertainties on the inherent tasks, most of the authors either assign them a given value or train them as additional hyperparameters~\cite{molnar_flow_2022, sun_physics-constrained_2020, linka_bayesian_2022}. This can lead to highly biased predictions, especially when setting the PDE-residual standard deviations $\sigma_f$ and $\sigma_h$ which introduce strong priors on the model adequacy. 

Recently Psaros et al. \cite{psaros_uncertainty_2023} discussed, \textit{inter alia}, alternatives generalizing the adjustment of some of these parameters — mainly the data-fitting standard deviations — in the context of unknown and heteroscedastic noise distributions. They either rely on \emph{offline learning} at the cost of a pre-trained Generative Adversarial Network (GAN) or \emph{online learning} of the weights based on additional parameter training. In particular, the number of these additional parameters may increase drastically when considering location-dependent variances, as suggested in \cite{psaros_uncertainty_2023}, for realistic applications and consequently suffer from computational costs. The open question remains on how to deal with such unknown (homo- or hetero-scedastic) noise distributions without adding computational complexity by learning additional hyperparameters.

Finally, although the question of physical model misspecification was pointed out in the total uncertainty quantification, the latter has not been addressed in \cite{psaros_uncertainty_2023} when misleading model uncertainty is assumed on the physical constraints $\mathcal{F}$ and $\mathcal{H}$. As a result, the issue of not introducing strong priors on the model adequacy by hand-tuning of the hyperparameters $\sigma_f$ and $\sigma_h$, usually unknown or prescribed, is still a challenging task. 

In view of this, we wanted to test the robustness of the usual BPINNs-HMC approach, as introduced in Sect. \ref{subsec:BPINNs_review}, on a test case demonstrating the issues arising from the multi-objective and multi-scale nature of the sampling using Sobolev training of neural networks.

\subsection{Sobolev training for BPINNs failure mode }
\label{subsec:Sobolev_failure}

\begin{figure}
    \centering
    \includegraphics[scale = 0.4]{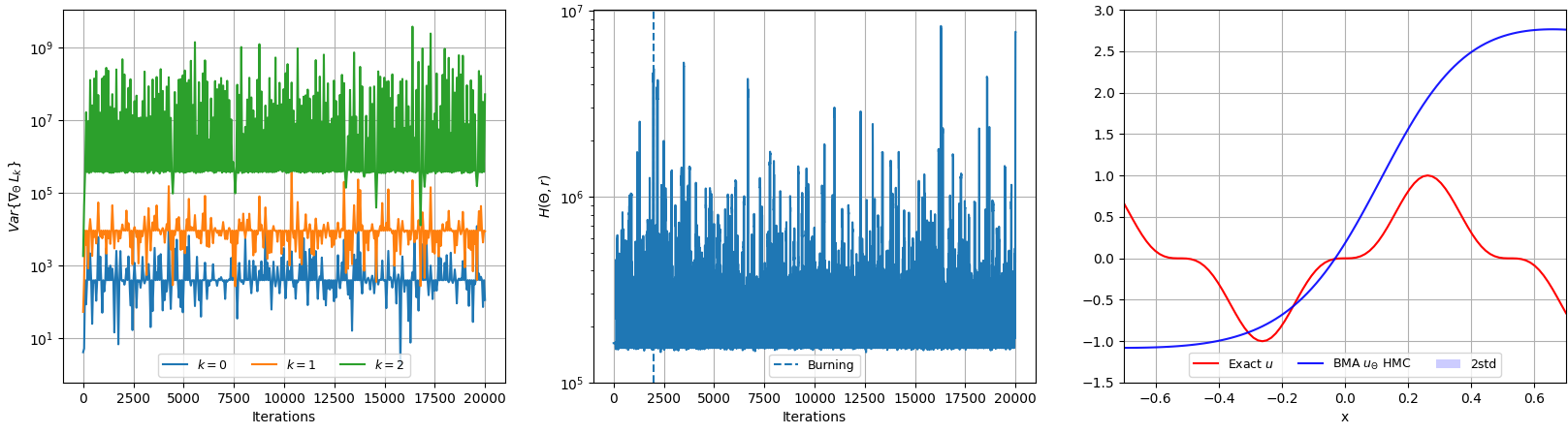}    
    \caption{ The HMC uniform-weighting failure mode for Sobolev training up to second-order derivatives leading to non-conservative Hamiltonian (on the middle), and extremely poor resulting approximation of the function (on the right). This is due to the strong imbalances in the variances of the effective gradient $\nabla_\Theta L_k$ distributions ($k=0,1,2$), plotted with respect to the ($N_s\times L$) HMC iterations (on the left).}
    \label{fig:2nd_order_uniform_w}
\end{figure}

Sobolev training is a special case of multi-objective BPINN sampling that likely leads to stiff learning due to the disparate scales involved \cite{maddu_inverse_2022}. Nevertheless, it is commonly used in the machine learning community to improve the prediction efficiency and generalization of neural networks, by adding information about the target function derivatives to the loss or its equivalent potential energy~\cite{czarnecki_sobolev_2017, Sob_training_PINNS, vlassis_sobolev_2021, avidan_sobolev_2022}.

This special training provides a baseline for testing the robustness of the present BPINNs-HMC method against the failure mode of vanishing task-specific gradients \cite{maddu_inverse_2022}. It also offers the opportunity to benchmark against the analytically $\varepsilon$-optimal weights that are known for Sobolev multi-objective optimization \cite{maddu_inverse_2022}. 

The BPINNs-HMC sampling is tested here on a Sobolev regression task, which means the dataset is restricted to $\mathcal{D} = \mathcal{D}^u$ involving measurements of a function and its derivatives $D^k_x$, $k\ge 1$  up to order $K$, such that the target posterior distribution is 
\begin{equation}
\label{Sop_post_dist}
    P(\Theta|\mathcal{D}) \propto \prod_{k=0}^K P(\mathcal{D}^u, D_x^k u|\Theta)P(\Theta)
\end{equation}
and the potential energy hence has the general form
\begin{equation}
\label{Sob_pot}
    \ds U(\Theta) = \sum_{k=0}^K \left[ \frac{\lambda_k}{2\sigma_k^2}\|D^k_x u_\Theta - D^k_x u\|^2 \right] +  \frac{\lambda_{K+1}}{2\sigma_{K+1}^2}\|\Theta\|^2 := \sum_{k=0}^{K+1} \lambda_k \mathcal{L}_k(\Theta)
\end{equation}
where $ L_k = \lambda_k\mathcal{L}_k$ refers to the weighted $k^{th}$ objective term, with $\lambda_k$ some positive weighting parameters to define (see Sect. \ref{subsec:InvDir_AW_HMC}). In this section, we use only a uniform weighting strategy, with $\lambda_k = 1, \forall k$, which corresponds to the classical BPINNs-HMC formulation. For sake of readability, equation \eqref{Sob_pot} gathers the log-prior terms of the neural network and inverse parameters, assuming they all have the same prior distribution. 

We first introduce a 1D Sobolev training up to second-order derivatives, with a test function $u(x) = \mathrm{sin}^3(\omega x)$ defined on $\Omega = [-0.7,0.7]$ for $\omega = 6$. We use 100 training points, set the leapfrog parameters $L=100$ and $\delta t = \num{1e-3}$ for the number of iterations and time step respectively, and perform $N_s = 200$ sampling iterations. We also restrict the test to a function approximation problem so that subsequently $\Theta$ refers only to the neural network parameters. In the following and unless otherwise indicated, all the $\sigma_k$ are equal to one since in practice we do not have access to the values of these parameters for the derivatives or residual PDE terms, but rather to the observation noise on the data field $u$ only, if available. 

Similarly to PINNs, this test case with uniform weights $\lambda_k$ leads to imbalanced gradient variances between the different objective terms. In particular, the higher-order derivatives present dominant gradient variances that contribute to the vanishing of the other tasks and lead to biased exploration of the posterior distribution. In Fig \ref{fig:2nd_order_uniform_w} (left) we see that the term $\mathrm{Var}\{ \nabla_\Theta L_2\}$ corresponding to the higher-order derivative quickly develops two orders of magnitude greater than the other effective gradient variances. In addition to inefficient exploration of the Pareto front, we also face instability issues, generated by the highest order derivative terms, that result in a lack of conservation of the Hamiltonian along the leapfrog trajectories (see Fig \ref{fig:2nd_order_uniform_w} middle). As specified in Sect. \ref{subsec:BPINNs_review}, such divergence pathologies on the classical HMC with uniform weighting are powerful diagnostics of bias in the resulting estimators and raise suspicions about the validity of the latter.  

\begin{figure}
    \centering
    \includegraphics[scale = 0.45]{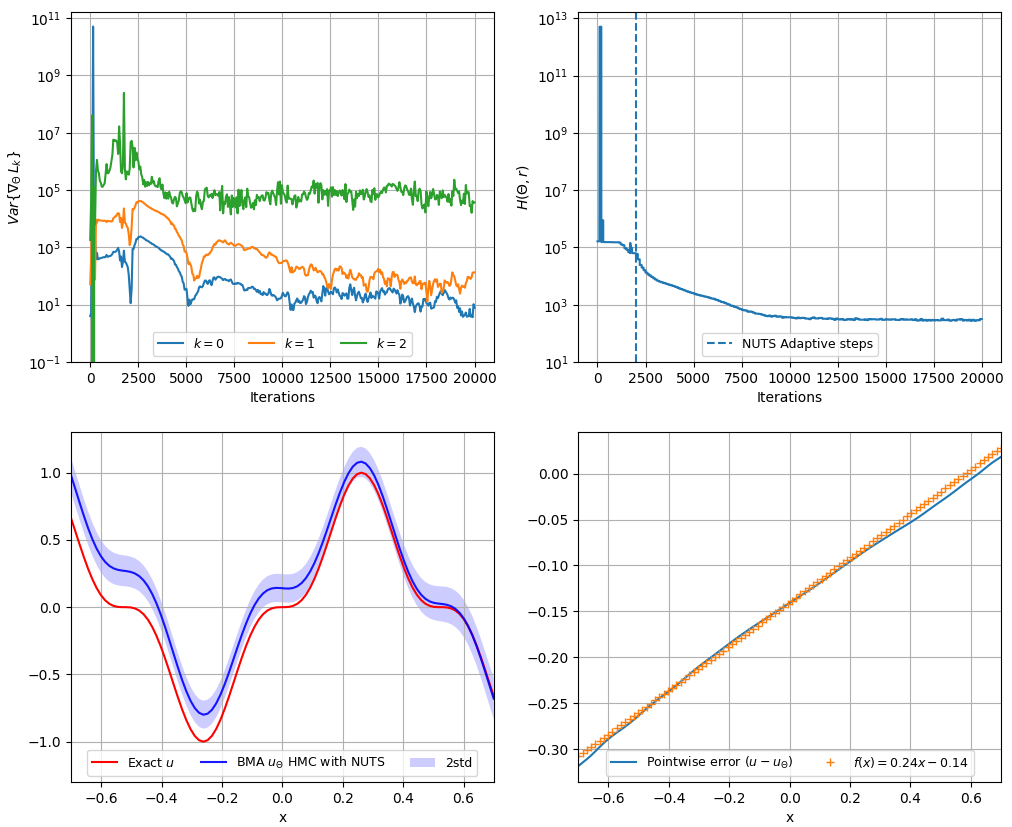}
    \caption{Failure mode of NUTS step-size adaptation in Sobolev training up to second order derivatives: variances of the effective gradient $\nabla_\Theta L_k$ distributions (top left) and Hamiltonian evolution (top right), respectively showing task imbalances and weak exploration of the energy levels. Signal approximation (bottom left) and pointwise error (bottom right) highlighting the linear deviation of $u_\Theta$. The vertical dotted line delimits the number of adaptive steps in the NUTS sampler. }
    \label{fig:2nd_order_nuts}
\end{figure}

An alternative to counteract these effects consists in reducing the time step $\delta t$ to balance the order of magnitude of the derivative terms and improve the Metropolis-Hasting acceptance rate of the BPINNs-HMC. However, a small time step within the leapfrog iterations is more likely to generate pathological random walk behaviors or biased sampling \cite{homan_no-u-turn_2014, betancourt_conceptual_2018}. To this aim, we attempt an adaptive strategy by using the No-U-Turn sampler (NUTS) with step-size adaptation, as detailed in Algorithm 5 from Hoffmann and Gelman \cite{homan_no-u-turn_2014} and implemented in the Python Open Source package hamiltorch \cite{cobb_scaling_2021}. We consider the same exact set of leapfrog parameters as previously — in order to comply with the same assumptions — and we impose $N = 20$ adaptive steps that lead to a final adapted time step of $\delta t = \num{1.29e-4}$. In this case, we again reached a configuration where we were not efficiently exploring the Pareto front, as evidenced by the variances of the effective gradients in Fig \ref{fig:2nd_order_nuts}. This resulted in a better approximation of the second derivative compared to the signal itself and demonstrated biased sampling in the sense that the signal $u$ is determined up to a linear function due to the prevalence of the higher derivative term. This linear deviation is also shown in Fig \ref{fig:2nd_order_nuts} — bottom right. This confirms that the NUTS time-step adaptation focuses rather on the prevailing conservation of the higher-order derivative which induced the stiffness. 

In short, even a simple 1D Sobolev training with trivial uniform weights induces major failure of the classical BPINNs-HMC approaches because of the sensitivity of the posterior distribution to the higher-order derivatives that generate instabilities. Consequently, such divergence in the Hamiltonian conservation renders the sampling approach inoperative. Moreover, the alternatives ensuring the Hamiltonian conservation are ineffective because they face either inefficient exploration of the energy levels or a strong imbalance in the multi-task and multi-scale sampling. This suggests that the Hamiltonian Markov chain cannot adequately explore the Pareto front of the target distribution resulting from this potential energy, and that strong imbalanced conditions cannot be overcome with the usual methodologies. 

The purpose is therefore to develop a strategy to provide balanced conditions between the different tasks, independently of their scales, by looking for an appropriate weighting formulation. This approach is essential regardless of the usual HMC concerns about the adaptive settings of the leapfrog parameters, and presents the advantage of reducing the instabilities without needlessly decreasing the time step.

\section{An Adaptive Weighting Strategy for Unbiased Uncertainty Quantification in BPINNs}
\label{sec:AW_HMC_for_BPINNs}

Conventional BPINN formulations exhibit limitations regarding multi-objective inferences, such as stability and stiffness issues, pathological conflicts between tasks, and biased exploration of the Pareto-optimal front. These problems cannot be tackled merely by adaptively setting the leapfrog parameters, as in the NUTS sampler, nor by hand-tuning the standard deviations $\sigma$, which introduces additional computational costs or energy waste. We therefore investigate another adaptive approach that focuses instead on the direct regularization of the target distribution: it aims to balance task weighting by automatically adapting the critical $\sigma$ parameters. 

\subsection{An Inverse Dirichlet Adaptive Weighting algorithm: AW-HMC}
\label{subsec:InvDir_AW_HMC}

The development of a new alternative considering the limits of the HMC-BPINN approach (previously discussed in Sect. \ref{sec:BPINNs_overview_limits}) becomes crucial, especially in the case of complex multi-objective problems arising from real-world data. This strategy must address the main pathologies identified by: 1) ensuring the exploration of the Pareto front of the target posterior distribution, 2) managing the scaling sensitivity of the different terms, and 3) controlling the Hamiltonian instabilities. 

Independently of these pathological considerations, there remains the issue of setting the critical $\sigma$ parameters, particularly when the level of noise on the data and the confidence in the PDE model are not prior knowledge. While manual tuning of these parameters is still commonplace, we could rely on the $\lambda$ weight adaptations to implicitly determine the noise and inherent task uncertainties rather than introduce strong priors on the model adequacy that may lead to misleading predictions.

In order to fulfill all these requirements, we consider an Inverse-Dirichlet based approach that has demonstrated its effectiveness in the PINNs framework when dealing with balanced training and multi-scale modeling \cite{maddu_inverse_2022}. It relies on adjusting the weights based on the variances of the loss term gradients, which can be interpreted as a training uncertainty with respect to the main descent direction in a high-dimensional multi-objective optimization problem. This strategy also offers considerable improvement in convergence over conventional training and avoids the vanishing of specific tasks. 

The idea of developing an Inverse Dirichlet adaptively weighted algorithm for BPINNs is to incorporate such training uncertainties on the different tasks within the Bayesian framework so that it can simultaneously take into account the noise, the model adequacy and the sensitivity of the tasks, all while ensuring Pareto front exploration. Therefore, we are trying to determine the positive weighting parameters $\lambda_k$, $k=0, ..., K$ in such a way that the weighted gradient $\nabla_\Theta L_k = \lambda_k \nabla_\Theta \mathcal{L}_k$ distributions of the potential energy terms have balanced variances. We propose to ensure gradient distributions with the same variance 
\begin{equation}
    \label{Inv_dir_var}
    \gamma^2 := \mathrm{Var}\{\lambda_k\nabla_\Theta \mathcal{L}_k \} \simeq \mathrm{min}_{t=0,...,K} (\mathrm{Var}\{\nabla_\Theta \mathcal{L}_t \}),\quad \forall k = 0,...,K 
\end{equation}
by setting the weights on an Inverse-Dirichlet based approach:
\begin{equation}
\label{Lambda_weights_Inv_Dir}
    \lambda_k =\left(\frac{\mathrm{min}_{t=0,...,K}(\mathrm{Var}\{\nabla_\Theta \mathcal{L}_t \})}{\mathrm{Var}\{\nabla_\Theta \mathcal{L}_k \} }\right)^{1/2} =\left( \frac{\gamma^2}{\mathrm{Var}\{\nabla_\Theta \mathcal{L}_k \} }\right)^{1/2}
\end{equation}
such that 
\begin{equation}
    \lambda_k \mathcal{N}(\mu_k, \mathrm{Var}\{ \nabla_\Theta \mathcal{L}_k\}) =  \left(\frac{\gamma^2}{\mathrm{Var}\{\nabla_\Theta \mathcal{L}_k \}}\right)^{1/2}\mathcal{N}(\mu_k, \mathrm{Var}\{ \nabla_\Theta \mathcal{L}_k\}) = \mathcal{N}(\mu_k, \gamma^2).
\end{equation} Note that we do not discuss here the case of $\lambda_{K+1}$ corresponding to the prior $P(\Theta)$, since the log-prior term acts rather as a $\L^2$-regularization in the equivalent constrained optimization problem, such that the weight balancing approach focuses essentially on the log-likelihood terms of the potential energy. In fact, the sampling should enable us to efficiently explore the Pareto front corresponding to balanced conditions between the data-fitting and the different PDE-based likelihood terms. On the contrary, we do not want to rely on a non-informative prior to achieve task balancing, so we impose the following upper bound
\begin{equation}
    \label{reg_prior}
    \mathrm{Var}\{\lambda_{K+1} \nabla_\Theta \mathcal{L}_{K+1} \} \le \gamma^2, 
\end{equation}
which can be achieved with setting $\lambda_{K+1} \le \sigma_{K+1}$, related to the assumption on the prior $P(\Theta)\sim\mathcal{N}(0,\sigma_{K+1}^2 I)$.
This comes from the observation that
\begin{equation}
    \lambda_{K+1}\nabla_\Theta \mathcal{L}_{K+1}(\Theta^{t_\tau}) =\frac{\lambda_{K+1}}{\sigma_{K+1}^2} \Theta^{t_\tau} \quad \text{s.t} \quad 
     \mathrm{Var}\{\lambda_{K+1} \nabla_\Theta \mathcal{L}_{K+1}(\Theta^{t_\tau}) \} =\frac{\lambda_{K+1}^2}{\sigma_{K+1}^4} \mathrm{Var}\{\Theta^{t_\tau}\} \leq \frac{1}{\sigma_{K+1}^2}\mathrm{Var}\{\Theta^{t_\tau}\}
\end{equation}
with $\Theta^{t_\tau}$ the set of parameters sampled at iteration $\tau$. The latter upper bound also provides a dispersion indicator between the posterior variance of $\Theta$ and its prior distribution, that can be used to set the value of $\sigma_{K+1}$ given $\gamma^2$.  

We investigate on-the-way methods to deal with the BPINNs-HMC failure mode, so that the weight adaptation strategy \eqref{Lambda_weights_Inv_Dir} depends on the sampling iterations $\tau$. This results in a modified Hamiltonian Monte Carlo, denoted Adaptively Weighted Hamiltonian Monte Carlo (AW-HMC) and detailed in Algorithm \ref{Adaptively Weighted Hamiltonian Monte Carlo}. The weighting strategy is carried on until a set number of adaptive iterations $N$, potentially different from the usual burning steps $M$. It assumes that $N\le M$, and enables us to reach a weighted posterior distribution, well-suited to the exploration of the Pareto front. In fact, finite adaptation preserves ergodicity and asymptotic convergence of the chain, while keeping $N\le M$ ensures the posterior distribution is drawn from the same weighted potential energy. In practice, the a priori burning phase is closely linked to the number of adaptive steps by taking $M = N$. We also introduce the notation $H_{\lambda_\tau}(\Theta, r)$ for the weighted Hamiltonian 
\begin{equation}
    \label{Weighted_Ham}
    H_{\lambda_\tau}(\Theta, r) = \sum_{k=0}^{K+1} \lambda_k(\tau) \mathcal{L}_k(\Theta) + \frac{1}{2}r^T\mathbf{M}^{-1}r
\end{equation}
which defines the new transition probability for the Metropolis-Hasting acceptance criterion. 

The present balancing of the target distribution, based on the minimum variance of the gradients \eqref{Lambda_weights_Inv_Dir} can be interpreted as adjusting the weights with respect to the most likely or the least sensitive term of the multi-potential energy. It therefore offers the advantage of improving the convergence of the BPINNs toward the Pareto-optimal solution and also enhances the reliability of the uncertainty quantification of the output, whose samples are drawn from the Pareto front. Indeed, this weighting strategy induces an automatic increase in the uncertainty of the least likely task by adaptively adjusting the $\lambda$ parameters. Such observations arise from the development of upper bounds for each of the gradient variances, as detailed in \ref{sec:bound_Inv_Dir}, which involves prediction errors and PDE residuals, as well as sensitivity terms characterizing the variability of the mean gradient descent directions for each task. In light of this, we were able to provide an upper bound on the joint variance $\gamma^2$ which is developed in equation \eqref{upper_bound_gamma} in a basic and general perspective.

Last but not least, the Inverse-Dirichlet based adaptive weighting relieves us from an unreasonable decrease in the time step, which no longer has to meet all the stiff scaling requirements to ensure Hamiltonian conservation. This approach then renders the sampling free of excessive tuning adaptation of the leapfrog hyperparameters $\delta t$ and $L$. In addition, this prevents pathological random-walk or divergence behaviors in the sampling since it enables the use of optimal integration time, both in terms of convergence rate and adequacy of the time step to the distinct learning tasks. 

The current AW-HMC algorithm is first validated on a Sobolev training benchmark with different complexities, which provides a basis for comparison with $\varepsilon$-optimality results. This also allows us to establish a new indicator for convergence diagnostics of the BPINNs. The robustness and efficiency of the present method are then experimented on more complex multi-task and multi-scale problems, along the different sections.

\RestyleAlgo{boxruled}
\begin{algorithm}
\label{Adaptively Weighted Hamiltonian Monte Carlo}
\caption{Adaptively Weighted Hamiltonian Monte Carlo (AW-HMC)}

\KwIn {Initial $\Theta^{t_0}$, $N_s$ number of samples, $L$ number of leapfrog steps, $\delta t$ leapfrog step size,\\ \qquad \quad $N$ number of adaptive iterations, $M$ burning steps and $\mathbf{M}$ the mass matrix.}
\BlankLine
\emph{Sampling procedure:} \\
\For {$\tau = 1...N_s$}{

Sample $r^{t_{\tau-1}} \sim \mathcal{N}(0, \mathbf{M})$\;
Set $(\Theta_0, r_0) \gets (\Theta^{t_{\tau-1}}, r^{t_{\tau-1}})$\;

\emph{Weights adaptation:} \\
    \eIf{$\tau \le N$}
    {\text{Compute }$\ds \lambda_k(\tau) =\left(\frac{\mathrm{min}_{j=0,...,K}(\mathrm{Var}\{\nabla_\Theta \mathcal{L}_j(\Theta_0) \})}{\mathrm{Var}\{\nabla_\Theta \mathcal{L}_k(\Theta_0) \} }\right)^{1/2} \quad \forall k = 0,...,K$ \text{and}  $\lambda_{K+1}(\tau) = 1$\;}
    {$\lambda_k(\tau) = \lambda_k(\tau-1) \quad \forall k = 0,...,K$ \text{and} $\lambda_{K+1}(\tau) = 1$}

\emph{Leapfrog:}\\
    \For{$i = 0...L-1$}
    {
    $\ds r_i \gets r_i - \frac{\delta t}{2} \sum_{k=0}^{K+1} \lambda_k(\tau) \nabla_\Theta \mathcal{L}_k(\Theta_i) $\;
    $\Theta_{i+1} \gets \Theta_i + \delta t \mathbf{M}^{-1}r_i$\;
    $\ds r_{i+1} \gets r_i -\frac{\delta t}{2} \sum_{k=0}^{K+1} \lambda_k(\tau) \nabla_\Theta \mathcal{L}_k(\Theta_{i+1}) $\; 
    }
\emph{Metropolis-Hastings:} \\
    Sample $p\sim \mathcal{U}(0,1)$\; 
    Compute $\alpha = \mathrm{min}(1, \mathrm{exp}(H_{\lambda_\tau}(\Theta_0 , r_0) - H_{\lambda_\tau}(\Theta_L, r_L) )$ using \eqref{Weighted_Ham}\;

    \eIf{$p\leq \alpha$}
    {$\Theta^{t_\tau} = \Theta_L$\;}
    {$\Theta^{t_\tau} = \Theta_0$\;}
    Collect the samples after burning : $\left\{\Theta^{t_i}\right\}_{i=M}^{N_s}$
}
\end{algorithm}

\subsection{Sobolev training benchmark and convergence diagnostics}
\label{subsec:Sob_train_bench}

\begin{figure}
    \centering
    \includegraphics[scale = 0.4]{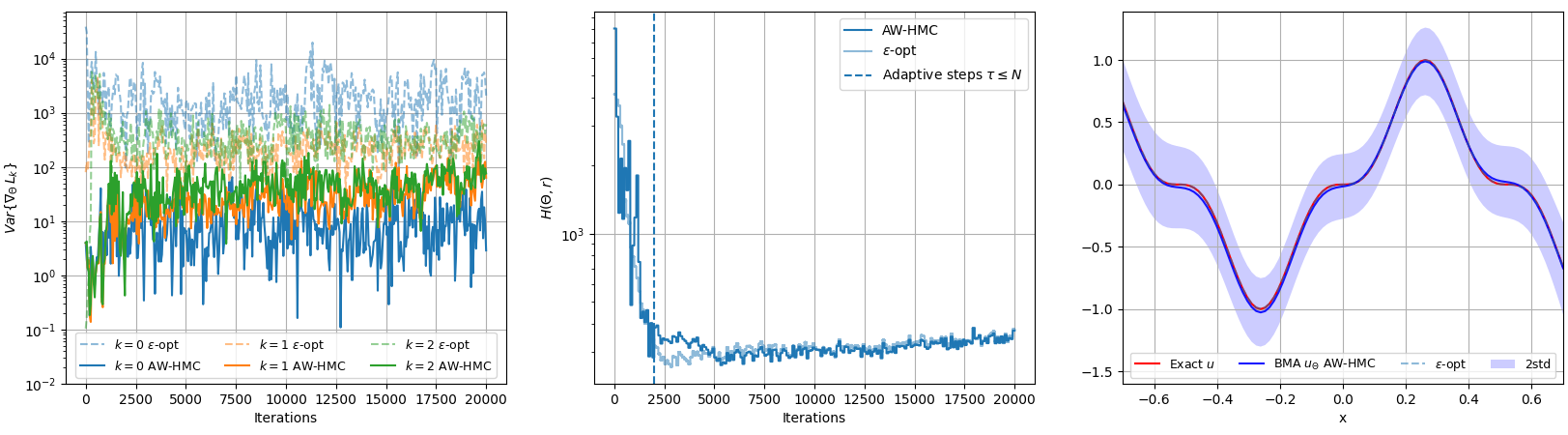}    
    \caption{Adaptively Weighted Hamiltonian Monte Carlo (AW-HMC) on Sobolev training up to second-order derivatives compared to $\varepsilon$-optimal weighting. Effective gradient distributions variances $\mathrm{Var}\{\nabla_\Theta L_k\}$ with balanced conditions between the tasks (on the left). Hamiltonian evolution throughout the sampling satisfying energy conservation (in the middle). Resulting field predictions with a comparison between $\varepsilon$-optimal results and AW-HMC strategy (on the right).}
    \label{fig:eps_aw_comp_k2}
\end{figure}

We investigate the performances of the proposed auto-weighted BPINN methodology on several applications, starting in this section with a Sobolev training benchmark. 
We first apply this new Adaptively Weighted strategy to the 1D Sobolev training introduced in Sect. \ref{subsec:Sobolev_failure} with the same exact set of hyperparameters. The number of adaptive steps is set to $N = 20$, as for the NUTS declination, to ensure an impartial comparison of the distinct methodologies.  

In addition, we compare the predictions with a reference case where the weights $\lambda_k$ are set accordingly to $\varepsilon$-optimal analytical solution \cite{Meer2022OptimallyWL}, which can be determined for Sobolev training based on equation \cite{maddu_inverse_2022}:  
\begin{equation}
  \lambda_k^\varepsilon = \frac{\prod_{j\neq k}\mathcal{I}_j}{\sum_{k=1}^K \prod_{j\neq k}\mathcal{I}_j} \quad \text{ s.t.~~}  \mathcal{I}_k = \int_\Omega |D_x^k(u)|^2 \,\mathrm{d}x.
\end{equation}
We tested our methodology against this $\varepsilon$-optimal solution assuming observation noise $\xi_u \sim \mathcal{N}(0, \sigma_u^2 I)$ such that $\sigma_k = \sigma_u, \forall k$ with $\sigma_u = 0.1$.
It provides good agreement between both approaches with a convergence of the Hamiltonian toward the same energy level, in Fig \ref{fig:eps_aw_comp_k2} (middle): in fact, the $\L^2$-relative error on the Hamiltonian values between the $\varepsilon$-optimal and AW-HMC methods scales around $\num{1e-4}$ after the adaptive steps. The AW-HMC method also provides $\L^2$-relative errors, compared to this optimal solution, ranging around $\num{1e-3}$ for both the signal and its derivatives. Finally, we also point out in Fig \ref{fig:eps_aw_comp_k2} (left) balanced gradient variances in the same way as observed with $\varepsilon$-optimal analytical weights. The AW-HMC methodology, therefore, provides similar results to $\varepsilon$-optimal solutions in terms of balance between the gradient variances, exploration of the Hamiltonian energy levels, and overall BMA predictions.   

Our new approach shows exceptionally balanced conditions between the different tasks: the effective gradient distribution variances $\mathrm{Var}\{\nabla_\Theta L_k\}$ present the same orders of magnitude throughout the training, even with a finite number of adaptive steps. This means that the posterior distribution reached after the auto-adjustment of the weights is well-suited to converging toward the Pareto front exploration, thus making the sampling more efficient. Preventing strong imbalance behavior on the gradient variances, and therefore task-specific bias has considerably improved the marginalization of such multi-objective potential energy, in comparison with the conventional approaches that presented major failures in Sect. \ref{subsec:Sobolev_failure}. 

\begin{figure}
    \centering
    \includegraphics[scale = 0.5]{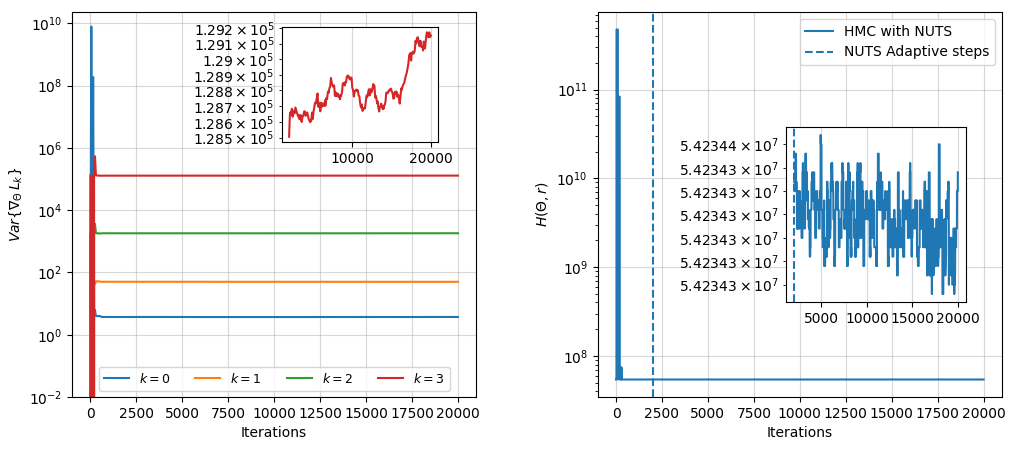}\\  
    \includegraphics[scale = 0.5]{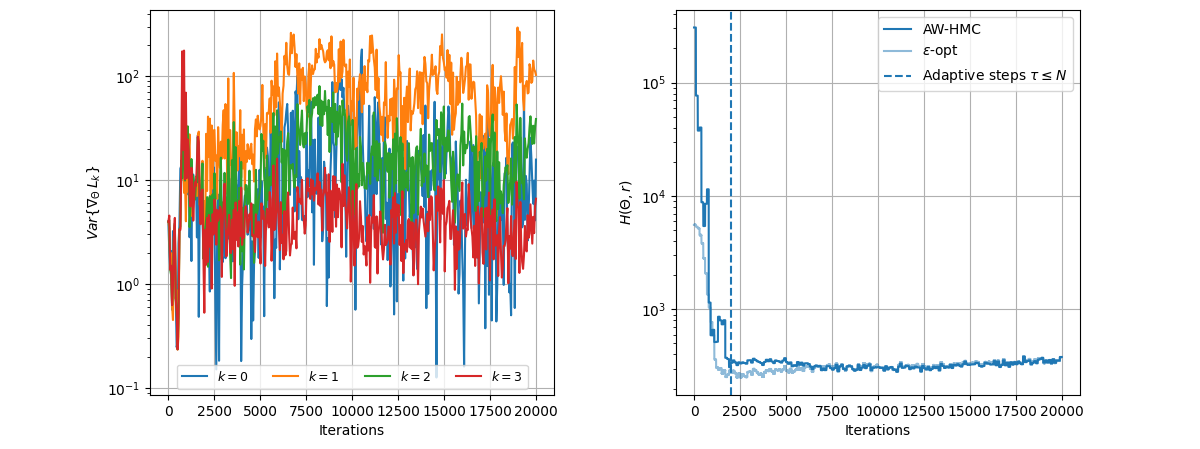}    
    \caption{Effective gradient distribution variances $\mathrm{Var}\{\nabla_\Theta L_k\}$ and Hamiltonian evolution on 1D Sobolev training up to the third-order derivative. The NUTS formulation (top) highlights strong imbalances between the learned tasks, on the left, and random-walk behaviors in exploring the energy level sets, on the right. AW-HMC strategy (bottom) and comparison with the $\varepsilon$-optimality. }
    \label{fig:eps_aw_comp_k3}
\end{figure}

\begin{figure}
    \centering
    \includegraphics[scale = 0.5]{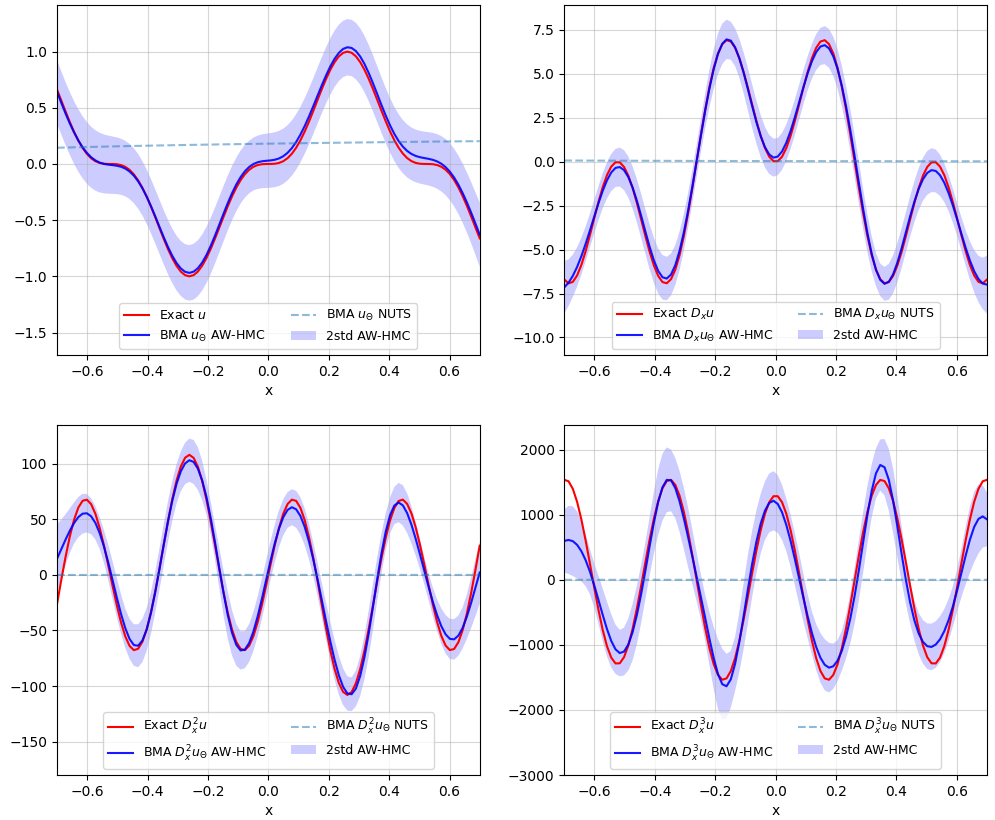}    
    \caption{1D Sobolev training up to third-order derivative: comparison of the BMA predictions, on the function and its derivatives, between the AW-HMC and NUTS formulations. The imbalance between tasks and random walk behavior of NUTS (see Fig \ref{fig:eps_aw_comp_k3}) results in ineffective BMA predictions. The AW-HMC methodology overcomes these effects and significantly improves the sampling of the target distribution.}
    \label{fig:nuts_aw_comp_k3}
\end{figure}

To further demonstrate the robustness of the method, we consider the third-order derivative extension of this test case, where even a NUTS adaptive strategy on the time step (reaching $\delta t = \num{1.36e-7}$) generates, here, pathological random-walk behavior making the sampling completely defective (Fig \ref{fig:eps_aw_comp_k3} top row and Fig \ref{fig:nuts_aw_comp_k3}). Such a significant decrease in the time step is clearly explained by the enhanced stiffness induced by the third-order derivative term in this multi-task learning. Indeed, the Hamiltonian trajectories are more likely to diverge during the deterministic steps due to this stiffness and require a small $\delta t$ to compensate for the divergence. To avoid the resulting pathological random walks, the overall integration time must be increased but this inevitably leads to excessive computational costs — under such a constraint on the leapfrog time step. This highlights the main limitation of NUTS when facing stiff multi-task sampling that involves separate scales.

\begin{figure}
    \centering
    \includegraphics[scale=0.55]{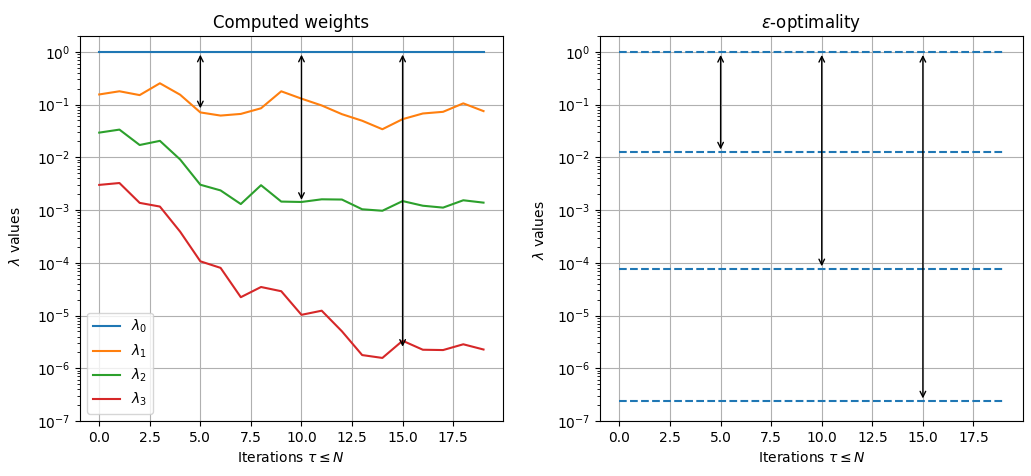}
    \caption{Evolution of the $\lambda_k$ weights along the adaptive steps ($\tau \le N$) on the left, and comparison with analytical $\varepsilon$-optimal weights for Sobolev training up to third-order derivative. Evolution of Average weights over several repetitions of the AW-HMC algorithm. This is induced by different initializations of the neural network parameters and momentum, to take into account 
    sample variability. The order of magnitude of the relative weights $\lambda_0/\lambda_i, \, i=1...3$ are represented by the double-headed arrows.}
    \label{fig:Weights_vs_Eps}
\end{figure}

In contrast, our approach overcomes these major failures (see Fig \ref{fig:nuts_aw_comp_k3}) without additional constraints on $\delta t$ and provides balanced gradient variances between the different tasks as illustrated in Fig \ref{fig:eps_aw_comp_k3} (bottom row). We also compare the results of the AW-HMC methodology with analytical weights from $\varepsilon$-optimality and show great agreement between the approaches. In addition, in order to deal with the stochastic-induced process of the BPINNs induced by sampling variability, we perform various repetitions of the sampling with different initialization of the neural network parameters and momentum. This leads to  averaged weight evolution along the adaptive steps presented in Fig \ref{fig:Weights_vs_Eps} that show the same order of magnitude as the analytical $\varepsilon$-optimal weights. 

\begin{figure}
    \centering
    \includegraphics[scale=0.51]{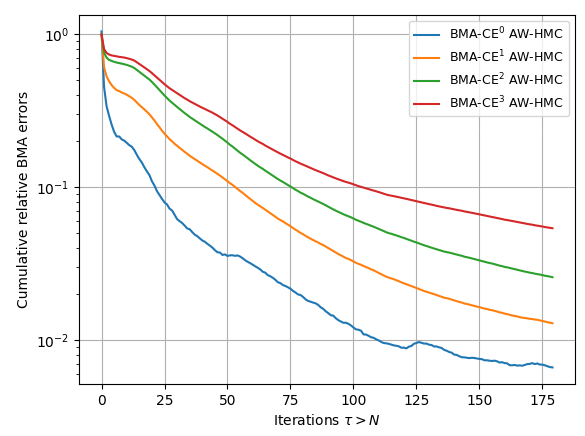}
    \hspace{1mm}
    \includegraphics[scale=0.53]{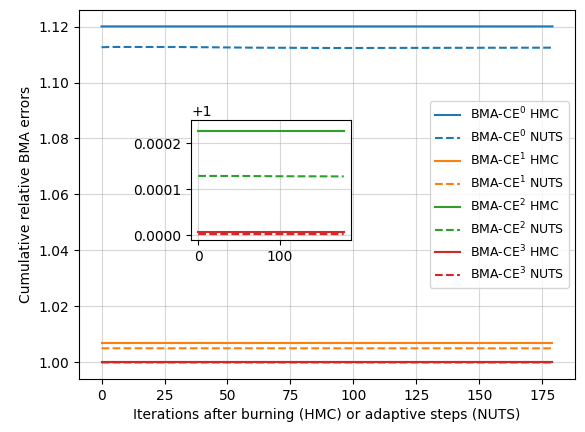}
    \caption{Cumulative relative BMA errors, computed according to equation \eqref{BMA_cum_error}, throughout the sampling iterations $\tau>N$ for Sobolev training up to third-order derivative. Comparison between the AW-HMC strategy (on the left) and classical HMC and NUTS formulations (on the right). These quantities remain nearly constant in pathological cases, due either to massive rejection or pathological random walk, highlighting the lack of convergence in the usual BPINNs-HMC formulations (on the right).}
    \label{fig:BMA_CE_1D}
\end{figure}

Apart from these qualitative comparisons between the different methodologies and the analytical solution, we subsequently introduce a new metric that quantifies the quality of the predictions. This complements the usual metrics with a convergence quantification of the sampling along the marginalization process. The samples collected after the burning steps in the AW-HMC process — i.e. all the instances of $\left\{\Theta^{t_i}\right\}_{i=M}^{N_s}$ — are first used to determine a Bayesian Model Average estimation as defined in equation \eqref{BMA}. Each sample provides a prediction $P(y|x, \Theta^{t_i} )$, for the neural network characterised by $\Theta^{t_i}$, and is theoretically drawn from the posterior distribution $P(\Theta |\mathcal{D}, \mathcal{M})$ such that the BMA is usually approximated by \cite{wilson_bayesian_2020}: 
\begin{equation}
\label{approx_BMA}
    P(y|x, \mathcal{D}, \mathcal{M}) \simeq \frac{1}{N_s-M}\sum_{i=M}^{N_s} P(y|x,\Theta^{t_i}) \quad \text{with} \quad \Theta^{t_i}\sim P(\Theta |\mathcal{D}, \mathcal{M}).
\end{equation}
In Sobolev training, we consider as the neural network outputs, the prediction of the function itself and all its derivatives $y = \left\{D_x^k u_\Theta,\ k = 0...K\right\}$, such that we can compute, according to equation \eqref{approx_BMA}, relative BMA errors with respect to each output defined by : 
\begin{equation}
    \label{BMA_rel_error}
    \ds \text{BMA-E}^k = \frac{\left\| P\left(D_x^k u_\Theta\,|\,x, \mathcal{D}, \mathcal{M}\right) -D_x^k u \right\|^2}{\|D_x^k u \|^2}, \qquad \forall k=0...K
\end{equation}
where the notation $\|\cdot\|$ used here refers to the functional $\L^2$-norm. Based on the previous definition and in order to incorporate convergence on the BMA along the marginalization process, we introduce a new diagnostic called cumulative (relative) BMA error, defined as follows: 
\begin{equation}
    \label{BMA_cum_error}
     \ds \text{BMA-CE}^k(\tau) = \frac{\left\|\ds \frac{1}{\tau-N}  \sum_{i=N}^\tau P\left(D_x^k u_\Theta\,|\,x, \Theta^{t_i}\right) -D_x^k u \right\|^2}{\left\|D_x^k u \right\|^2}, \qquad \forall k=0...K
\end{equation}
depending on the sampling iterations after the adaptive steps, for $\tau > N$ in Algorithm \ref{Adaptively Weighted Hamiltonian Monte Carlo}. These formulae can be directly extended to all the neural network outputs, in a more general framework and quantify the sampling efficiency in terms of convergence rate. The cumulative BMA errors are represented in Fig \ref{fig:BMA_CE_1D} for the third-order extension of Sobolev training highlighting the convergence of the AW-HMC sampler for each of the functional tasks (on the left). Instead, these quantities remain nearly constant for the pathological HMC and NUTS formulations, due to massive rejections and random-walk behavior, respectively (see Fig \ref{fig:BMA_CE_1D} on the right).  

We finally extended this Sobolev training test case to several benchmarks on 2D, where we studied the impact of the functional complexity and the number of training points on the Bayesian Model Average errors. The details of these benchmark problems and the training setup are provided in \ref{sec:2D_Sob_training}
and have shown enhanced robustness and efficiency of the AW-HMC algorithm for the BPINNs. 

\subsection{A multi-scale Lokta-Volterra inverse problem}
\label{subsec:Lokta_Volterra}

\begin{figure}
    \centering
    \includegraphics[scale = 0.5]{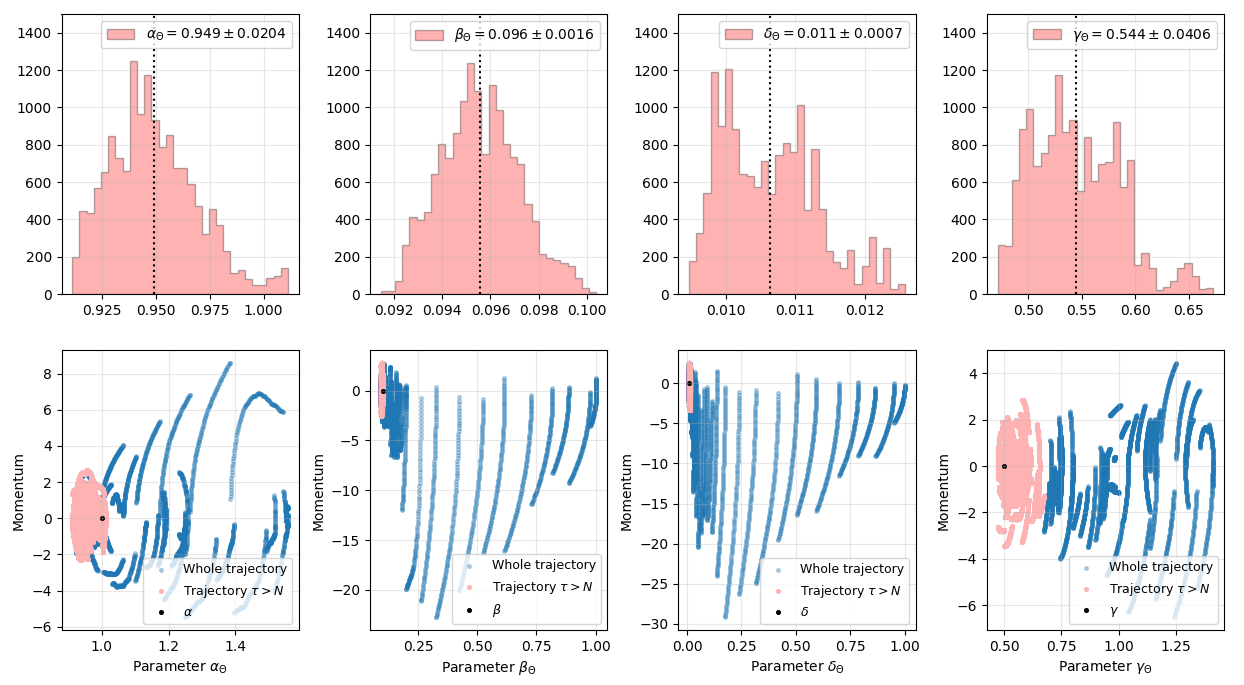}
    \caption{Lokta-Volterra multi-scale inference: histogram of the marginal posterior distributions for the inverse parameters $\alpha_\Theta,\, \beta_\Theta,\, \delta_\Theta$ and $\gamma_\Theta$ (top). Phase diagrams of the parameter trajectories throughout the sampling (bottom) that characterized convergence toward their respective modes during the adaptive steps (in blue) and efficient exploration of the mode neighborhood after the adaptive steps (in red). The ground truth parameters are respectively $\alpha = 1,\, \beta = 0.1,\, \delta = 0.01$ and $\gamma = 0.5$ and establish an inverse problem with separate scales.}
    \label{fig:LV_inv_params}
\end{figure}

\begin{figure}
    \centering
    \includegraphics[scale = 0.7]{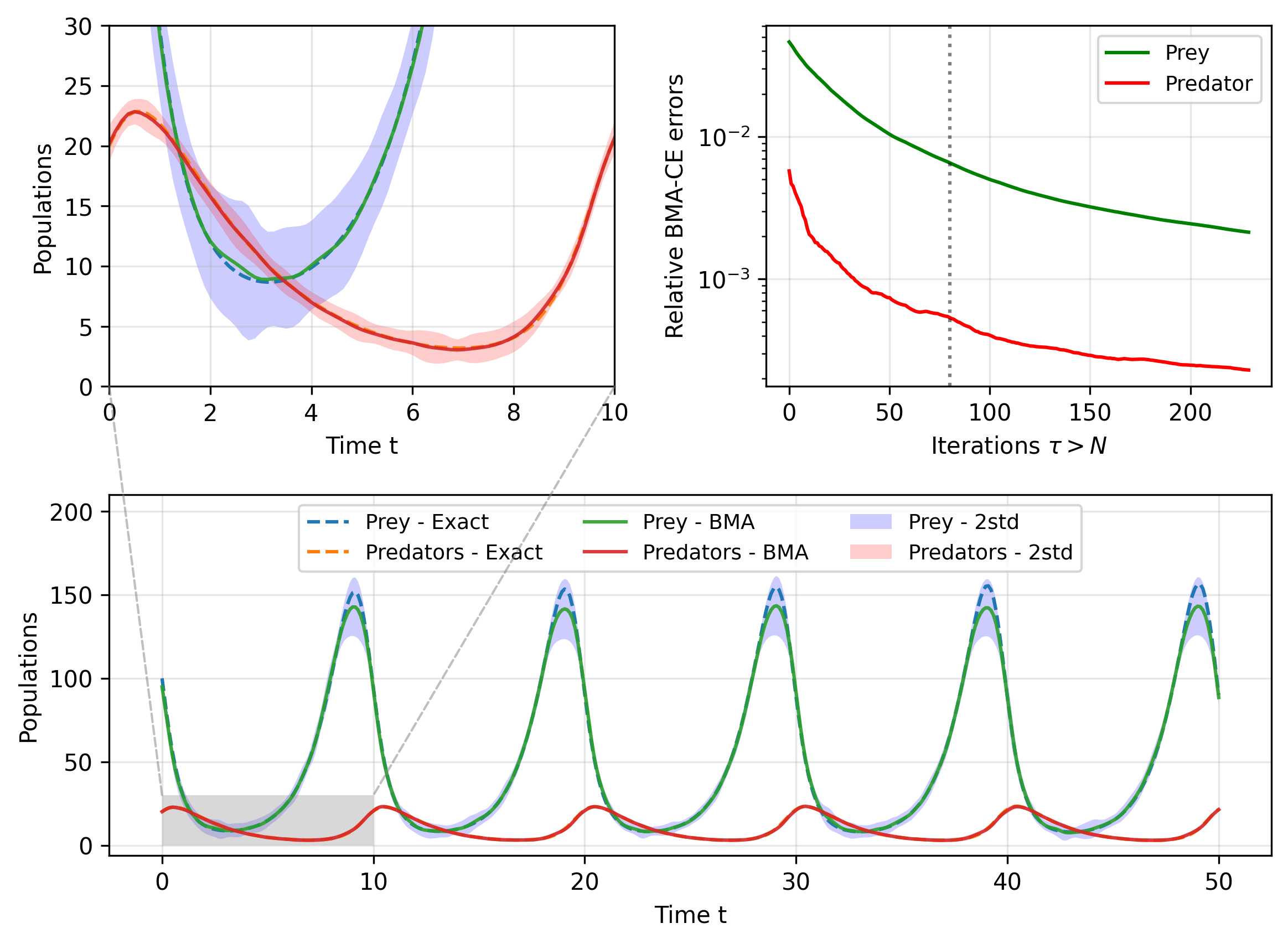}
    \caption{Lokta-Volterra multi-scale inference: BMA predictions of the two-species populations along the physical time with their uncertainties, bottom and top left. Relative BMA-CE errors defined as in equation \eqref{BMA_cum_error} for the neural network outputs $y=(u_\Theta,v_\Theta)$ plotted throughout the sampling iterations, — top right. The dotted vertical line marks the introduction of the ODE-likelihood terms in the sequential training.}
    \label{fig:LV_BMA}
\end{figure}

We demonstrate the use of AW-HMC on a multi-scale dynamical inverse problem to quantify the impact of the scaling. As Linka et al. \cite{linka_bayesian_2022} pointed out, sensitivity to scaling that may hinder the performance of classical BPINNs, especially when considering nonlinear dynamical systems. The multi-scale nature and stiffness resulting from real-world problems, where vanishing task-specific gradients are commonplace, is therefore an interesting benchmark to quantify the robustness of the present method.

In this context, we consider a Lokta-Volterra dynamical system with parameters of highly varying orders of magnitude defined by the following ordinary differential equation (ODE) system: 
\begin{equation}
    \label{LV_ODE}
    \left\{\begin{array}{l}
    \ds \frac{\mathrm{d} u}{\mathrm{d}t}= \alpha u -\beta uv, \quad t\in \Omega\\[3mm]
    \ds \frac{\mathrm{d} v}{\mathrm{d}t}= \delta uv -\gamma v, \quad t \in\Omega\\[2mm]
    u(0) = u_0,\ v(0) = v_0
    \end{array}\right.
\end{equation}
which characterizes the temporal evolution of predator-prey species. The notations $u(t)$ and $v(t)$ respectively refer to the prey and predator population size at a given time $t$, whereas the parameters $\alpha, \beta, \delta, \gamma \ge 0$ control the population dynamics, as growing and shrinkage rates. Thereafter, we set the initial populations to $u_0 = 100$ and $v_0 = 20$ with the following parameters $\alpha = 1,\, \beta = 0.1,\, \delta = 0.01$ and $\gamma = 0.5$ intentionally selected with different orders of magnitude. This sets up an inverse problem benchmark based on real-world dynamics with separate scales involved.

The observation data are first numerically generated by solving the ODE system \eqref{LV_ODE} on a uniform temporal grid $\Omega = [0,50]$ with a thin resolution of $400$ points. The data are randomly sampled so as to consider only half in the training phase of the different samplers. The dataset $\mathcal{D}$ then involves these partial measurements of $u$ and $v$ at $200$ different times, potentially with some added noise, and the same collocation points are kept to satisfy the ODE constraints. In this section, we focus on an inverse problem by inferring the unknown model parameters $\Sigma = \left\{ \alpha, \beta, \delta, \gamma \right\}$ from these measurement data while recovering the whole species evolution on the original finer resolution. 

Regarding the noticeable scaling difference between the two populations, we consider a predator-prey split of the tasks such that each field $u$ and $v$ satisfies a data-fitting likelihood term and an ODE-residual likelihood term. We also assume log-normal prior distributions on $\Sigma$ to ensure positivity of the inverse parameters, as Yang et al. \cite{yang_b-pinns_2021} have shown that such priors improve the inference, and we set independent normal distributions on the neural network parameters $\theta$. In practice though, we use a change of variable by introducing $\Sigma = e^{\tilde{\Sigma} }:= \left\{e^{\tilde{\alpha}}, e^{\tilde{\beta}}, e^{\tilde{\delta}}, e^{\tilde{\gamma}} \right\}$ for each of the inverse parameters to infer $\tilde{\Sigma} $ assuming normal prior distributions as well. For this test case, we impose weakly informed priors, especially on $\tilde{\Sigma}$, since we expect our methodology to handle the multi-scale inference due to the unbiased auto-weighting of the tasks. We therefore assume that both the neural network and inverse parameters all gather the same prior distribution, given by $\Theta\sim\mathcal{N}(0,\sigma_\Theta^2 I_{p+d})$ where $\Theta =\left\{\theta, \tilde{\Sigma} \right\}$.
 
Under these assumptions, we can define the multi-potential energy of the corresponding Hamiltonian system: 
\begin{equation}
    \label{LV_pot_energy}
    \begin{aligned}
    U(\Theta) &= \frac{\lambda_0}{2\sigma_0^2}\left\|u_\Theta - u\right\|_{\mathcal{D}}^2
    +  \frac{\lambda_1}{2\sigma_1^2}\left\|v_\Theta - v\right\|_{\mathcal{D}}^2
    + \frac{\lambda_2}{2\sigma_2^2}\left\|\frac{\mathrm{d}u_\Theta}{\mathrm{d}t} - \alpha_\Theta u_\Theta + \beta_\Theta u_\Theta v_\Theta \right\|_{\mathcal{D}}^2 \\
     & \qquad + \frac{\lambda_3}{2\sigma_3^2}\left\|\frac{\mathrm{d}v_\Theta}{\mathrm{d}t} - \delta_\Theta u_\Theta v_\Theta + \gamma_\Theta v_\Theta \right\|_{\mathcal{D}}^2
     + \frac{1}{2\sigma_\Theta^2} \|\Theta\|^2_{\R^{p+d}}
     \end{aligned}
\end{equation}
where the inferred inverse parameters are defined by $\Sigma_\Theta = e^{\tilde{\Sigma_\Theta}}$ and we also set all the $\sigma_\bullet$ equal to one, as we do not wish to impose strong priors on the tasks and model uncertainty. As mentioned previously in Sect. \ref{subsec:MO_paradigm}, the norms are respectively the RMS and the Euclidean norm for the last term. The prior on the parameters is assumed to follow a Gaussian distribution with a larger standard deviation $\sigma_\Theta = 10$, in the sense that a slightly diffuse distribution induces weakly informed priors on the $\Theta$ parameters. This also ensures that constraint \eqref{reg_prior} for a non-informative prior is satisfied.

\begin{table}[t]
        \renewcommand{\arraystretch}{1.2}\centering
    \begin{tabular}{|c|wc{1.5cm}|wc{1.5cm}|wc{1.5cm}|wc{1.5cm}|}
        \hline
        \diagbox{Seq. step}{$\lambda_k$} & $\lambda_0$ & $\lambda_1$   & $\lambda_2$& $\lambda_3$ \\ \hline
        Data-fitting (step 1) & $\num{3.83e-2}$  & 1  &  —  & —   \\ \hline
        Data-fitting + ODE tasks (step 2) & $\num{4.87e-2}$   &  1  & $\num{9.16e-3}$  & $\num{1.16e-1}$  \\ \hline
        \multicolumn{5}{c}{ }\\[-3mm] \hline
        \diagbox{Seq. step}{$\tilde{\sigma}_k$} & $\tilde{\sigma}_0$  & $\tilde{\sigma}_1$  & $\tilde{\sigma}_2$  & $\tilde{\sigma}_3$   \\ \hline
        Data-fitting (step 1) & 5.109   & 1  &  —  & —   \\ \hline
        Data-fitting + ODE tasks (step 2) & 4.531  &  1  & 10.45  & 2.936  \\ \hline
    \end{tabular}
    \caption{Weight parameters $\lambda_k$ obtained after the adaptive steps in the Lokta-Volterra multi-scale inverse problem, for each of the sequential steps (top rows). Effective standard deviations $\tilde{\sigma}_k$ resulting from the weight adaptations and computed as $\ds \tilde{\sigma}_k = \sqrt{1/\lambda_k}$ for each of the sequential steps (bottom rows). This highlights enhanced uncertainties on the tasks related to the prey species. The splitting of the sequential steps is detailed in Sect. \ref{subsec:Lokta_Volterra}.}
    \label{tab:LV_lambdas}
\end{table}

For such inverse modeling, the sampling is decomposed using sequential training. This means that 1) the neural network parameters are sampled with an AW-HMC strategy to mainly target the data-fitting likelihood terms (setting $\lambda_2 = \lambda_3 = 0$).  2) We then introduce the ODE-residual tasks in \eqref{LV_pot_energy} to provide estimations of the missing inverse parameters, using the AW-HMC algorithm with initial neural network parameters $\theta^{t_0}$ resulting from 1). The BMA predictions and uncertainty quantification finally rely on this entire sampling procedure. In the two-step sequential training, the number of adaptive and sampling iterations are first set to $N = 20$ and $N_s = 100$, and then $N=50$ and $N_s=200$ while the leapfrog parameters are given by $L=100$, $\delta t = \num{5e-4}$ and $\num{2e-4}$ respectively, for the time steps in 1) and 2). The neural network itself is composed of 4 layers with 32 neurons per layer and we use the sin activation function considering the periodic nature of the solution for the Lokta-Volterra system. 

On such an inverse problem the classical BPINNs-HMC algorithm faces massive rejection because the Hamiltonian trajectories are not conserved, which results in  inoperative sampling (Fig \ref{fig:LV_HMC_Fail}). Even the adaptive strategies on the time step struggle to deal with the multi-scale dynamics and require an extreme decrease in the $\delta t$  value to obtain some stability, as detailed in \ref{sec:LV_fails}. The natural implication of such constraints on the leapfrog time step is lack of convergence toward the Pareto front and poor inference of the inverse parameters, subject to weakly informed priors (see Fig \ref{fig:LV_HMC_NUTS_Fail} and \ref{fig:LV_inv_params_Fail} from \ref{sec:LV_fails}). In fact, Linka et al. \cite{linka_bayesian_2022} addressed the same issue on learning COVID-19 dynamics and imposed (in Sect. 4.3 of \cite{linka_bayesian_2022}) log-normal prior distributions on the inverse parameters that already rely on appropriate scaling. The need for such appropriate scaling strongly impacts the inference in the sense that it requires prior knowledge which biases the sampling.

On the contrary, we assume independent priors with respect to the scaling and show that our approach is able to properly recover all the $\Sigma$ parameters as well as predict the species evolution with minimal tuning and decrease on $\delta t$. The recovery of separate scales no longer requires prior knowledge of the inverse parameter scaling to converge to their respective modes. The results shown in Fig \ref{fig:LV_inv_params} represent both the marginal posterior distributions of each inferred inverse parameter $\Sigma_\Theta$ and their trajectories when exploring the phase space distribution $\pi(\Theta,r)$. For the latter, we plotted the entire sampling trajectories that converge toward their respective mode during the adaptive steps, to finally sample around them as illustrated by the final trajectories for $\tau>N$. This confirms the ability of AW-HMC to quickly identify the separate modes of this inverse problem and manage such multi-scale dynamics. 

In order to quantify the effectiveness in identifying the parameters, we also measure the relative error in the inference of the parameters $\Sigma_\Theta = \{\alpha_\Theta, \beta_\Theta, \delta_\Theta, \gamma_\Theta\}$
\begin{equation}
    \label{LV_param_err}
    E_{\Sigma_\Theta} = \frac{|\Sigma_\Theta - \Sigma|}{\Sigma}
\end{equation}
where the prediction is given by $\ds \Sigma_\Theta = \frac{1}{N_s - N}\sum_{i= N}^{N_s} e^{\tilde{\Sigma_\Theta}^{t_i}}$, and we show that these relative errors all scale around $\num{5e-2}$ for the four inverse parameters. The predictive evolution of the species populations is displayed in Fig \ref{fig:LV_BMA}, as a BMA on the neural network outputs $y=(u_\Theta, v_\Theta)$, and compared to the exact solutions in a qualitative and quantitative way. In this sense, we computed relative BMA cumulative errors for both the species, highlighting the convergence of the sampling, top right of Fig \ref{fig:LV_BMA}. We see that the insertion of the ODE-residual likelihood terms in the two-step sequential training improves the convergence of the predictions when compared to pure data-based sampling. 

This test case also reveals higher uncertainties on the evolution of the prey population characterized by effective standard deviations about four times greater (see Table \ref{tab:LV_lambdas}). The enhanced uncertainty on these specific tasks is highlighted by smaller values of $\lambda_0$ and $\lambda_2$ at the end of the adaptive steps, compared to $\lambda_1$ and $\lambda_3$ in the potential energy \eqref{LV_pot_energy}. Therefore, the AW-HMC strategy benefits from its ability to adaptively weight the $\lambda$ parameters to intrinsically characterize the task uncertainties based on their gradient variances.

\section{Application to Computational Fluid Dynamics: Stenotic Blood Flow}
\label{sec:Results}

We illustrate the use of the methodology set out in Sect. \ref{subsec:InvDir_AW_HMC} in a real-world problem from fluid mechanics, more precisely the study of inpainting and inverse problems on incompressible stenotic flows in asymmetric geometries. The objective is to demonstrate the generalization and performance of the present AW-HMC algorithm on more complex 2D geometries and nonlinear PDE dynamics under noise and sparsity of the data.

The measurement data are generated by randomly sampling the fully resolved Computational Fluid Dynamics (CFD) solutions on scattered locations. The direct numerical simulation of vascular flows in asymmetric stenotic vascular geometries is performed using a meshless solver based on the Discretization-Corrected Particle Strength Exchange (DC PSE) method as detailed in \cite{bourantas_using_2016}.

\subsection{Inpainting problem with sparse and noisy data}
\label{subsec:Inpainting_pb}

\begin{figure}
    \hspace{-10mm}
    \includegraphics[scale = 0.7]{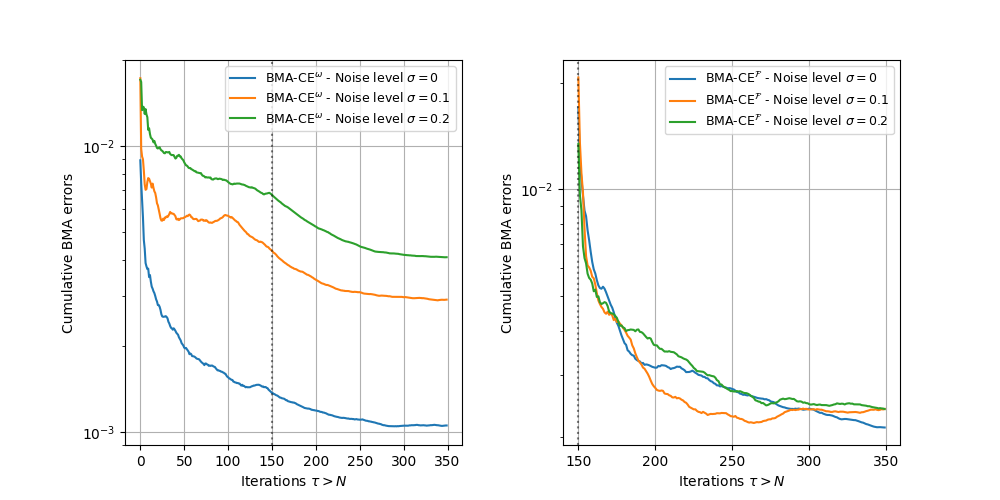}
    \caption{Vorticity physics-informed inpainting problem: Bayesian Model Average Cumulative Error diagnostics, as defined in \eqref{BMA_CE_inpainting}, throughout the sampling iterations and for different noise levels. BMA-CE on the vorticity field prediction (on the left) and on the PDE residual $\mathcal{F}$ satisfying \eqref{vort_NS} (on the right). The dotted vertical lines mark the introduction of the PDE constraint in sequential training. }
    \label{fig:noise_comp_vorticity}
\end{figure}

\begin{figure}
    \centering
    \includegraphics[scale = 0.55]{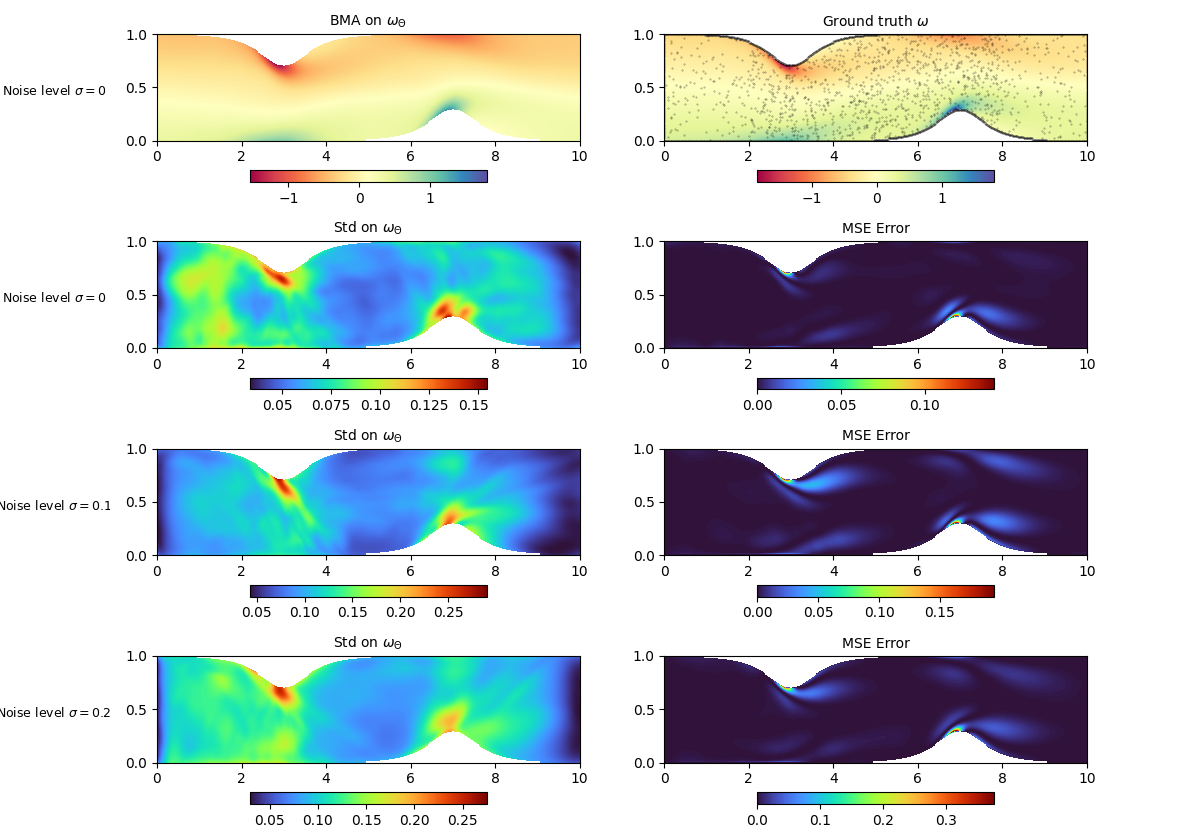}
    \caption{Physics-informed inpainting problem: BMA prediction of the vorticity field $\omega_\Theta$ in asymmetric stenosis without noise, compared to the ground truth solution $\omega$ (top). The black dots on the exact field correspond to the training measurements of the dataset $\mathcal{D}$.   
    Comparison of the uncertainty standard deviations (Std) and mean squared errors (MSE) on the predicted vorticity field $\omega_\Theta$ for different noise levels ($\sigma=0, 0.1, 0.2$), shown in the bottom rows. }
    \label{fig:noise_comp_uncertainty}
\end{figure}

Inpainting problems have drawn increasing interest in MRI or CT medical imaging as an opportunity to reduce artifacts and recover missing information by using deep learning approaches \cite{inpainting_2020_MRI, Inpainting_Gan_2020, multi_task_inpainting_2021}. Although the usual inpainting framework incorporates only measurement data in the image processing, Zheng et al. investigated a physics-informed version of the problem by incorporating the underlying physics as indirect measurements \cite{zheng_physics-informed_2020_inpainting}. The present section falls within the same context — the idea is to infer the whole flow reconstruction based on sparse and noisy measurements while imposing PDE constraints on some complementary collocation points. 

The governing equations of the stenotic flow dynamic are written here in a velocity $\mathbf{u}=(u,v)$ and vorticity $\omega$ formulation in two dimensions, satisfying an incompressible steady-state Navier-Stokes equation given by  
\begin{equation}
    \label{vort_NS_0}
    \ds (\mathbf{u}\cdot \nabla)\mathbf{\omega}= {Re}^{-1} \Delta \omega, \quad \text{in } \Omega\\
\end{equation}
or equivalently
\begin{equation}
    \label{vort_NS}
    u\frac{\partial \omega}{\partial x}+v\frac{\partial \omega}{\partial y} = \frac{1}{Re} \Delta \omega, \quad \text{in } \Omega
\end{equation}
where $Re$ refers to the dimensionless Reynolds number, $\omega$ is the vorticity field $\ds \omega = \frac{\partial v}{\partial x} -\frac{\partial u}{\partial y}$ and the incompressibility condition ensures $\nabla \cdot \mathbf{u} = 0$. 
We consider the 2D stenotic spatial domain $\Omega \subset [0,10]\times[0,1]$ and assume two different kinds of boundary conditions: 1) the stenosis upper and lower walls, denoted $\partial \Omega_1$, where we impose no-slip conditions such that $\mathbf{u}_{\partial\Omega_1} = 0$ and $\omega = (\nabla \times \mathbf{u})_{\partial\Omega_1}$ and 2) the inlet and outlet boundaries, denoted $\partial\Omega_2$, with a prescribed parabolic profile and Neumann condition, respectively, on the velocity in the main flow direction. These boundary conditions are detailed in Sect. 4.4 of the DC PSE article \cite{bourantas_using_2016}. We also first consider that the Reynolds number is known and set to $Re = 200$ according to the CFD simulations, such that the set of parameters to infer $\Theta$ is restricted here to the neural network weights and bias. 

\begin{table}[t]
    \centering\renewcommand{\arraystretch}{1.2}
    \begin{tabular}{|c|wc{1.5cm}|wc{1.5cm}|wc{1.5cm}|wc{1.5cm}|}
        \hline
        \diagbox{Noise level}{$\lambda_k$} & $\lambda_0$ & $\lambda_1$   & $\lambda_2$   & $\lambda_3$   \\ \hline
        $\sigma = 0$ & 1 & 0.46  &  0.88  & 0.51   \\ \hline
        $\sigma = 0.1$ & 1   &  0.19  & 0.35  & 0.29  \\ \hline
        $\sigma = 0.2$ & 1   &  0.12  & 0.39  & 0.27  \\ \hline
                \multicolumn{5}{c}{ }\\[-3mm] \hline
         \diagbox{Noise level}{$\tilde{\sigma}_k$} & $\tilde{\sigma}_0$  & $\tilde{\sigma}_1$  & $\tilde{\sigma}_2$  & $\tilde{\sigma}_3$   \\ \hline
        $\sigma = 0$ & 1 & 1.47  &  1.07  & 1.40   \\ \hline
        $\sigma = 0.1$ & 1   & 2.29   & 1.69 & 1.86  \\ \hline
        $\sigma = 0.2$ & 1   &  2.89  &  1.60 & 1.92  \\ \hline
    \end{tabular}
    \caption{Final $\lambda_k$ weight parameters for each $\sigma$ noise level in the physics-informed inpainting problem (top rows). Effective $\tilde{\sigma}_k$ standard deviations resulting from the weight adaptations and computed as $\ds \tilde{\sigma}_k = \sqrt{1/\lambda_k}$ for each noise level (bottom rows). This highlights the overall adaptation of the effective standard deviations to the noise magnitude and the task sensitivities to the noise level. In particular, the wall-boundary conditions associated with $\lambda_1$ present the highest noise sensitivity.}
    \label{tab:inpainting_lambdas}
\end{table}

The measurement dataset, $\mathcal{D}$, is composed of noisy vorticity data on $\mathcal{D}^{\partial_1}$ and $\mathcal{D}^{\partial_2}$, defined as in \eqref{D_bc} respectively for sets $\partial\Omega_1$ and $\partial\Omega_2$, as well as on 1282 interior collocation points $\mathcal{D}^\omega$ that cover less than 2\% of all the data required for the full vorticity field reconstruction on $\Omega$. We finally defined the $\mathcal{D}^\Omega$ dataset as 6408 interior points representing 6\% of the entire reconstructed data field, where we require that the PDE \eqref{vort_NS} be satisfied in a physically-constrained inpainting formulation. The multi-potential energy is then defined by:
\begin{equation}
\label{pot_vort}
    \begin{aligned}
    U(\Theta) &= \frac{\lambda_0}{2\sigma_0^2}\left\|\omega_\Theta - \omega \right\|_{\mathcal{D}^{\omega}}^2
    + \frac{\lambda_1}{2\sigma_1^2}\left\|\omega_\Theta -\restriction{\omega}{ \partial\Omega_1}
 \right\|_{\mathcal{D}^{\partial_1}}^2
    +  \frac{\lambda_2}{2\sigma_2^2}\left\|\omega_\Theta -\restriction{\omega}{ \partial\Omega_2}
 \right\|_{\mathcal{D}^{\partial_2}}^2  \\
    & \qquad + \frac{\lambda_3}{2\sigma_3^2}\left\|  u_n \frac{\partial \omega_\Theta}{\partial x} + v_n \frac{\partial \omega_\Theta}{\partial y} - \frac{1}{Re} \Delta \omega_\Theta \right\|_{\mathcal{D}^{\Omega}}^2
     + \frac{1}{2\sigma_\Theta^2} \|\Theta\|^2_{R_{p}}
     \end{aligned}
\end{equation}
with $u_n$ and $v_n$ noisy evaluations of the velocity field on the $\mathcal{D}^\Omega$ set. We then used sequential training by adding the PDE-residual likelihood term in the second sampling phase, such that the AW-HMC parameters are given first by $N=50$ and $N_s=200$, and then $N=50$ and $N_s=250$ for a leapfrog path length $L=150$ and time step $\delta t = \num{5e-4}$. As for the previous benchmarks, we set all the $\sigma_\bullet$ equal to one and assume a centered normal distribution with the standard deviation $\sigma_\Theta = 10$ for the neural network parameters prior. The neural network is composed of 4 layers with 32 neurons per layer and is based on the hyperbolic tangent activation function. The velocity and vorticity CFD solutions $(\mathbf{u}, \omega)$ are both corrupted by additive Gaussian noise such that $\bullet_n = \bullet+\sigma\xi$, where $\xi\sim\mathcal{N}(0,\psi^2)$ is a vector of element-wise independent and identically-distributed Gaussian random numbers with mean zero and variance $\psi^2 = \mathrm{Var}\{\bullet\}$, and $\sigma$ refers to the level of added noise. 

In this physics-informed inpainting problem, we investigate the impact of the level of noise $\sigma$ on the BMA predictions of the vorticity field, as well as on the physical constraint by extending the notion of BMA convergence to the PDE residual. Hence, we compute the BMA-CE diagnostics for the field $\omega$ and the PDE constraint based on
\begin{equation}
\label{BMA_CE_inpainting}
    \begin{aligned}
        &\ds \text{BMA-CE}^\omega(\tau) = \left\|\ds \frac{1}{\tau-N}  \sum_{i=N}^\tau P\left(\omega_\Theta\,|\,x, \Theta^{t_i}\right) -\omega \right\|^2\\
        &\ds \text{BMA-CE}^{\mathcal{F}}(\tau) = \left\|\ds \frac{1}{\tau-N}  \sum_{i=N}^\tau P\left(\mathcal{F}(\omega_\Theta) \,|\,x, \Theta^{t_i}\right) \right\|^2
    \end{aligned}    
\end{equation}
with $\mathcal{F}(\omega_\Theta)$ the evaluation of the PDE from equation \eqref{vort_NS}. The comparative curves for different noise levels are represented in Fig \ref{fig:noise_comp_vorticity} and show sampling convergence toward final BMA errors that scale about $\num{1.05e-3}$, $\num{2.9e-3}$ and $\num{4.08e-3}$, respectively, for noise levels $\sigma=0,\ 0.1$ and $0.2$. In addition, we see that the PDE residual constraints converge independently to the noise level, reaching final BMA errors around $\num{2e-3}$ in all cases. 

To supplement the performance quantification of the inpainting formulation in recovering the entire vorticity field along with its uncertainty, we also use the Prediction Interval Coverage Probability (PICP) metric as defined by Yao et al. \cite{Yao2019QualityUQ}. This consists of a quality indicator of the posterior approximation, which evaluates the percentage of the ground truth observations contained within 95\% of the prediction interval, as given by: 
\begin{equation}
    \label{PICP}
        PICP = \frac{1}{N} \sum_{i=1}^N \mathbb{1}_{(\omega_\Theta^l)_i \,\le\, \omega_i \,\le\, (\omega_\Theta^h)_i}
\end{equation} where $\omega_\Theta^l$ and $\omega_\Theta^h$ are respectively the 2.5\% and 97.5\% percentiles of the predictive distribution on the vorticity. In this case, the notation $N$ refers to the total number of observations in the predictive dataset, in other words, the grid resolution of the computational domain $\Omega$. 
In our application, this PICP metric shows that more than 99\% of the vorticity ground truth observations are covered by the posterior distribution of the neural network output $\omega_\Theta$, independently of the level of noise. 

We also expect our self-weighted adaptation of $\lambda_k$ to be able to capture noise sensitivity with respect to the value of $\sigma$, and intrinsic task sensitivities to noise level without imposing any a priori on the noise level estimation. This is the key point of our methodology since we intentionally decouple $\sigma_k$ in \eqref{pot_vort} from the noise magnitude, and rely on the self-weighted strategy to quantify their related uncertainties. On the contrary, when dealing with noisy measurement data in applications researchers frequently assume the fidelity of each sensor to be known and set the standard deviations $\sigma_k$ accordingly. They can also be defined as additional learnable parameters to be inferred. The latter is usually subject to additional computational costs in \emph{online} learning or requires alternative neural network formalism used as pre-training in \emph{offline} learning \cite{psaros_uncertainty_2023}. In contrast, the strength of the AW-HMC methodology relies on its similar computational cost compared to classical BPINNs-HMC. Moreover, AW-HMC improves convergence by drawing attention to exploring the Pareto front with optimal integration time. Therefore, it can shorten overall sampling requirements making this a competitive strategy in terms of computational cost.

The results presented in Fig \ref{fig:noise_comp_uncertainty} demonstrate the noise resistance of the AW-HMC approach and highlight sensitivity consideration with respect to the noise and tasks (see Table \ref{tab:inpainting_lambdas}). We first noticed differences in the auto-adjustment of the lambda values relative to noise levels, leading to global enhanced uncertainties with increasing noise. We also observed various uncertainty adjustments depending on the sensitivity of the different tasks to the noise. In fact, the comparison of the local standard deviations on the vorticity field in Fig \ref{fig:noise_comp_uncertainty} shows that the wall boundary conditions are the most sensitive to noise, automatically increasing the uncertainties in these areas. The inlet and outlet boundaries are rather less sensitive. This is highlighted by a lower adaptation of their uncertainties to the noise level. In short, this application has shown the ability of our new adaptive methodology to automatically adjust the weights, and with them the uncertainties, to the intrinsic task sensitivities to the noise and to adapt the uncertainty to noise magnitude itself.

\subsection{Inverse problem with parameter estimation and latent field recovery}
\label{subsec:Inv_pb}

\begin{figure}
    \hspace{-5mm}
    \includegraphics[scale = 0.47]{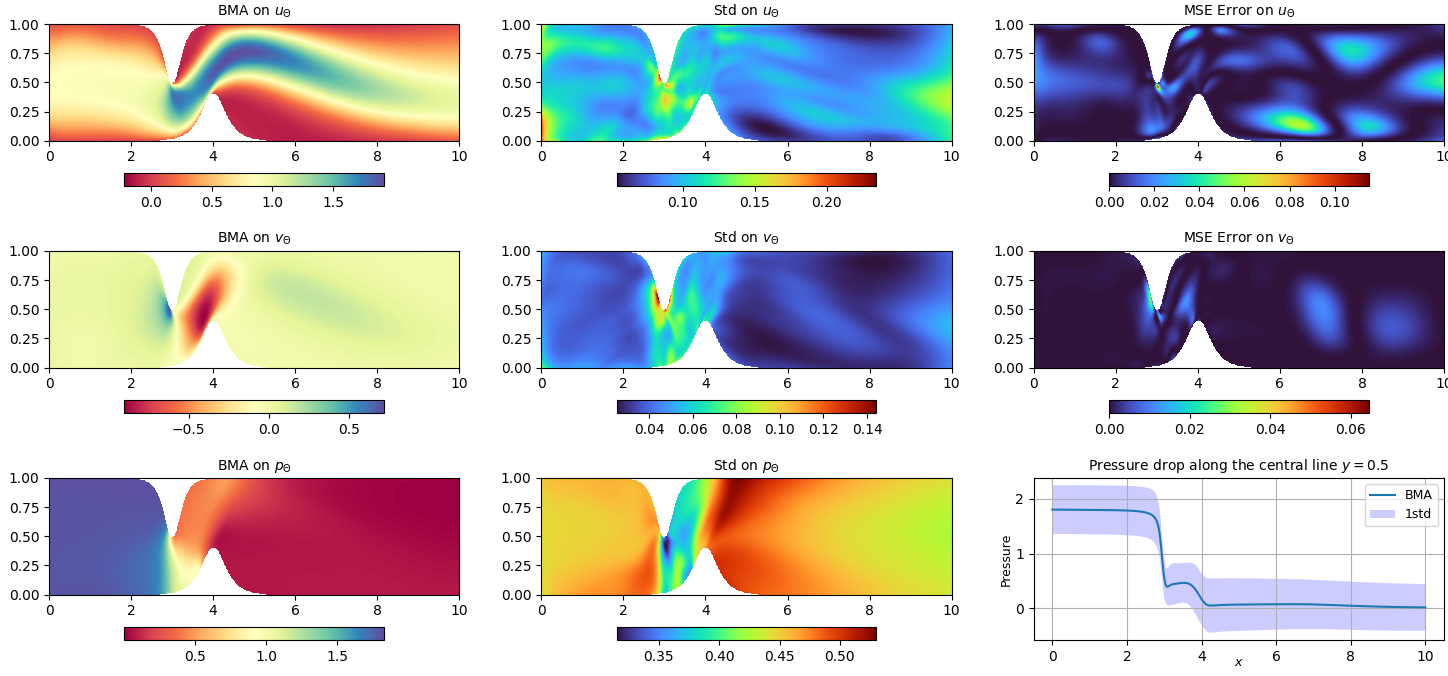}
    \caption{CFD inverse problem: BMA predictions of the velocity field $\mathbf{u}_\Theta = (u_\Theta, v_\Theta)$ in asymmetric stenosis along with their uncertainty standard deviations (Std) and mean squared errors (MSE), at the top. BMA and uncertainty on the inferred latent pressure field with the pressure evolution plotted along the central line $y=0.5$, leading to an average pressure drop of $1.78$ —  bottom.}
    \label{fig:velocity_BMA}
\end{figure}

\begin{figure}
    \centering
    \includegraphics[scale = 0.7]{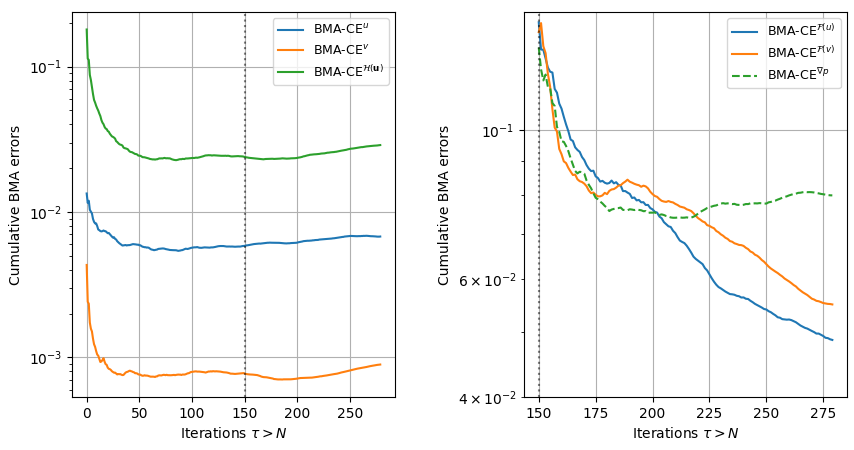}
    \caption{CFD inverse problem: Bayesian Model Average Cumulative Errors throughout the sampling iterations for the velocity field components $\mathbf{u}=(u,v)$, the divergence-free condition $\mathcal{H}(\mathbf{u})$, on the left, and the PDE residuals $\mathcal{F}(u)$ and $\mathcal{F}(v)$, on the right. The dotted curve represents the a posteriori checking of pressure gradient norm BMA-CE error as defined in equation \eqref{BMA_CE_grad_p}}
    \label{fig:BMA_CE_velocity}
\end{figure}

\begin{figure}
    \centering
    \includegraphics[scale = 0.5]{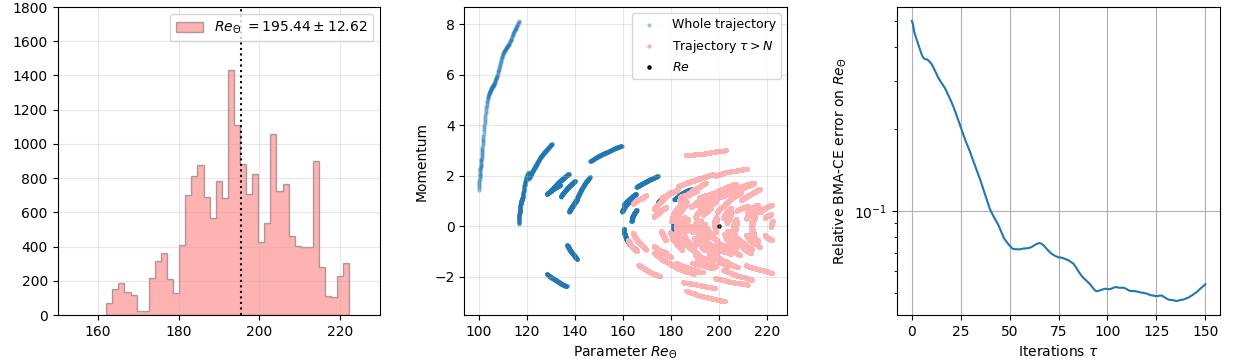}
    \caption{CFD inverse problem: from left to right, histogram of the marginal posterior distribution for the inverse Reynolds parameter, phase diagram of its trajectory throughout the sampling and BMA-CE error using the absolute relative norm as defined in \eqref{BMA_CE_Re}. The relative BMA-CE error on $Re_\Theta$ is plotted over the all the $\tau$ iterations of the second step sampling in the sequential training. }
    \label{fig:Reynolds_inv}
\end{figure}

As a second CFD application, we consider a multi-objective flow inverse problem in an asymmetric and steep stenosis geometry. This aims to provide both a parameter estimation of the flow regime and recover a hidden field using our adaptively weighted strategy. Such considerations, motivated by real-world applications, use incomplete or corrupted measurement data in an attempt to derive additional information, which remains challenging or impractical to obtain straightforwardly. 

With an emphasis on physical and biomedical problems, Raissi et al. investigated the extraction of hidden fluid mechanics quantities of interest from flow visualizations, using physics-informed deep learning \cite{Raissi2018HiddenFM, raissi_hidden_2020}. The authors relied only on measurements of a passive scalar concentration that satisfied the incompressible Navier-Stokes equations, to infer the velocity and pressure fields in both external and internal flows. 

In this direction, we focus on the velocity $\mathbf{u}=(u,v)$ and pressure $p$ formulation of the stenotic flow dynamics such that the continuity and momentum governing steady-state equations are written: 
\begin{equation}
    \label{NS_velocity}
    \left\{\begin{aligned}
    \ds (\mathbf{u}\cdot \nabla)\mathbf{u}&= - \nabla p + {Re}^{-1} \Delta \mathbf{u}, &\quad \text{in } \Omega\\
    \ds \nabla \cdot \mathbf{u} &= 0, &\quad \text{in } \Omega
    \end{aligned}\right.
\end{equation}
under the incompressibility condition on the stenotic domain $\Omega\subset[0,10]\times[0,1]$. We impose adherent boundary conditions on the wall interfaces such that $\mathbf{u}_{\partial\Omega_1} = 0$, and the following inlet/outlet boundary conditions respectively:
\begin{equation}
    \begin{aligned}
         u = 4y-4y^2,\, v = 0 \quad &\forall (x,y)\in\{0\}\times[0,1]\\
         \frac{\partial u}{\partial x} = 0, \, v = 0 \quad &\forall (x,y)\in\{10\}\times[0,1].
    \end{aligned}
\end{equation}
The direct numerical simulation is performed using the DC-PSE formulation \cite{bourantas_using_2016} with a Reynolds number set to $Re=200$, as in the previous section. It is used to generate the observation data on $\Omega$ with a thin resolution. The $\mathcal{D}$ dataset is then composed of partial measurements of $\mathbf{u}$ randomly sampled to consider 9559 training points, representing less than 3\% of the entire target resolution. The same collocation points are included to impose the PDE constraints, denoted $\mathcal{F}(\mathbf{u}):= (\mathcal{F}(u), \mathcal{F}(v)) $, as well as the diverge-free condition $\mathcal{H}(\mathbf{u})$. 

Finally, we set up the inverse problem by inferring the flow regime, considering the Reynolds number as an unknown model parameter $\Sigma = \{Re\}$. At the same time, we address the multi-task problem to recover the latent pressure from the partial measurements of the velocity field and the fluid flow dynamics assumptions. The pressure field prediction, in particular, is adjusted throughout the sampling in such a way that its gradient satisfies the governing equations \eqref{NS_velocity}. As commonly established by the nature of the Navier-Stokes equation, the pressure is though not uniquely defined and, given the lack of precise boundary conditions on this field, is thus determined up to a constant. The predictions of each of the quantities of interest, namely the velocity and pressure, are then recovered on the original finer resolution in Fig \ref{fig:velocity_BMA}. As in Sect. \ref{subsec:Lokta_Volterra}, we select a log-normal prior distribution for the physical parameter and independent normal distributions for the neural network parameters, and we also use a sequential training approach, incorporating the PDE constraints in the second sampling phase.

The validation of the inference is first performed by computing the BMA-CE diagnostics for the velocity field components, the PDE constraints, and the incompressibility condition written in the same way as in equation \eqref{BMA_CE_inpainting}. The results are provided in Fig \ref{fig:BMA_CE_velocity} and highlight the convergence of each term toward final BMA errors scaling respectively about $\text{BMA-CE}^u(N_s)=\num{6.4e-3}$, $\text{BMA-CE}^v(N_s)=\num{1e-3}$, $\text{BMA-CE}^{\mathcal{F}(\mathbf{u})}(N_s)=(\num{4.2e-2},\, \num{4.7e-2})$ and $\text{BMA-CE}^{\mathcal{H}(\mathbf{u})}(N_s)=\num{2.9e-2}$. The Bayesian Model Average predictions of the velocity field are then compared in Fig \ref{fig:velocity_BMA} with the ground truth observations providing local mean squared error (MSE) that are embedded in their uncertainties and show enhanced standard deviations at the regions with higher errors. The PICP metric also enables to estimate that more than $95\%$ of the velocity field ground truth is recovered by the posterior distribution of $\mathbf{u}_\Theta$.

For the inverse parameter, we computed a BMA cumulative error based on the relative L1-norm defined as follows
\begin{equation}
    \label{BMA_CE_Re}
     \text{BMA-CE}^{Re}(\tau) =  \frac{ \left|\ds \frac{1}{\tau}  \sum_{i=1}^\tau Re_{\Theta^{t_i}} -Re \right|}{|Re|}, \qquad \forall \tau=1...N_s
\end{equation}
where $Re_{\Theta^{t_i}}$ refers to the prediction of the Reynolds number for the sample characterized by the parameters $\Theta^{t_i}$. We show in Fig \ref{fig:Reynolds_inv} that this relative error converges, reaching at the end of the sampling a residual of $\num{5.4e-2}$. We also represent here the histogram of the marginal posterior distribution of $Re_{\Theta}$ and its trajectory in the phase space illustrating the convergence toward its mode during the adaptive steps $\tau < N$. In fact, our approach leads to an estimate of the Reynolds number, inferred  from the measurements data $\mathcal{D}$, which is consistent with the exact value and results in the predictive interval $Re_\Theta\in \left[182.82, 208.06\right]$.  

The latent pressure field BMA, inferred up to constant, is illustrated in Fig \ref{fig:velocity_BMA} with its uncertainty and is able to capture a sharp pressure drop —estimated in average to $1.78$ — arising from the steep stenosis geometry. In fact, it has been emphasized by Sun et al. in symmetric geometries, that such pressure drops turn to become nonlinear as the stenotic geometry becomes narrower \cite{sun_surrogate_2020}, which is in line with what we obtain in our asymmetric case. As the pressure ground truth is unknown in this application, we complement the validation of the inverse problem with a-posteriori checking on the pressure gradient. In this sense, we provide a PICP estimate on the pressure recovery which stands around $91\%$ for its gradient norm, but also introduce the following posterior diagnostic on the pressure BMA-CE error:
\begin{equation}
\label{BMA_CE_grad_p}
    \ds \text{BMA-CE}^{\nabla p}(\tau) = \left\|\, \ds \frac{1}{\tau-N}  \sum_{i=N}^\tau \left |P\left(\nabla p_\Theta\,|\,x, \Theta^{t_i}\right)\right| -\left| \nabla p \right| \, \right\|^2\\
\end{equation}
where $\left| \,\cdot\, \right|$ denotes the vector norm, and $\nabla p$ is the evaluation of the exact gradient pressure from equation \eqref{NS_velocity}. The results are plotted, in dotted line, throughout the sampling iterations in Fig \ref{fig:BMA_CE_velocity}, and reach a residual error of $\num{7.6e-2}$. This illustrates good agreement between the ground truth and the predictive pressure gradient arising from our adaptively-weighted strategy. 

Overall, the present AW-HMC methodology relies on multi-task sampling to identify the flow regime through partial  measurements of the velocity field and thus handles a complex flow inverse problem with latent field recovery that satisfies non-linear physical PDE constraints.

\section{Concluding remarks}
\label{sec:Conclusion}

BPINNs have recently emerged as a promising deep-learning framework for data assimilation and a valuable tool for uncertainty quantification (UQ) \cite{yang_b-pinns_2021}. This offers the opportunity to merge the predictive power of Physics-Informed Neural Networks (PINN) with UQ in a Bayesian inference framework using Markov Chain Monte Carlo (MCMC) sampling. This makes it possible to quantify the confidence in predictions under sparse and noisy data with physical model constraints, which is especially appealing for applications in complex systems. For this, Hamiltonian Monte Carlo has been established as a powerful MCMC sampler due to its ability to efficiently explore high-dimensional target distributions \cite{betancourt_conceptual_2018}. With it, BPINNs have extended the use of PINNs to a Bayesian UQ setting. 

As we have shown here, BPINNs, however, share similar failure modes as PINNs: the multi-objective cost function translates to a multi-potential sampling problem in a  BPINN. This presents the same difficulties in balancing the inference tasks and efficiently exploring the Pareto front as found in standard PINNs \cite{pareto_pinns}. We illustrated this in a  Sobolev training benchmark, which is prone to stiffness, disparate scales, and vanishing task-specific gradients. We emphasized that BPINNs are sensitive to the choice of the $\lambda$ weights in the potential energy, which can possibly lead to biased predictions or inoperative sampling. Hence, the standard weighting strategy appears to be inefficient in multi-scale problems and multi-task inference, while it turns out to be unsustainable to manually tune the weights in a reproducible and reliable way. Recently proposed alternatives \cite{psaros_uncertainty_2023} are subject to additional hyper-parameter tuning or pre-training of the weights with a GAN, at the expense of increased computational complexity. Also, previous approaches mainly focused on measurement noise estimation and did not include physical model mis-specification concerns which are also critical, especially when UQ modeling is the goal. 

Robust automatic weighting strategies are therefore essential to apply BPINNs to multi-scale and multi-task (inverse) problems and improve the reliability of the UQ estimates. Here, we have therefore proposed the AW-HMC BPINN formulation, which provides a plug-in automatic adaptive weighting strategy for standard BPINNs. AW-HMC effectively deals with multi-potential sampling, energy conservation instabilities, disparate scales, and noise in the data, as we have shown in the presented benchmarks.

We have shown that the presented strategy ensures a weighted posterior distribution well-fitted to explore the Pareto front, providing balanced sampling by ensuring appropriate adjustment of the $\lambda$ weights based on Inverse Dirichlet weighting \cite{maddu_inverse_2022}. The weights can therefore directly be interpreted as training uncertainties, as measured by the variances of the task-specific training gradients. This leads to weights that are adjusted with respect to the model to yield the least sensitive multi-potential energy for BPINN HMC sampling. This results in  improved convergence, robustness, and UQ reliability, as the sampling focuses on the Pareto front. This enables BPINNs to effectively and efficiently address multi-task UQ.

The proposed method is also computationally more efficient than previous approaches, since it does not require additional hyper-parameters or network layers. This also ensures optimal integration time and convergence in the leapfrog training. This prevents time steps from tending to zero or becoming very small, avoiding a problem commonly encountered in No-U-Turn Sampling (NUTS) when attempting to avoid the pathologically divergent trajectories characteristic of HMC instabilities. The present methodology improves the situation, since the time step no longer needs to meet all of the stiff scaling requirements to ensure energy conservation. As a result, it shortens overall integration time and sample number requirements, combining computational efficiency with robustness against sampling instabilities. 

Our results also show that AW-HMC reduces bias in the sampling, since it is able to automatically adjust the $\lambda$ parameters, and with them the uncertainty estimates, according to the sensitivity of each term to the noise or inherent scaling. In classical approaches, this is prohibited by the bias and implicit prior introduced by manual weight tuning. In fact, we demonstrated the efficiency of the present method in capturing inverse parameters of different orders of magnitude in a multi-scale problem, assuming completely independent priors with respect to the scaling. Previously, this would have been addressed by imposing prior distributions on these parameters that already rely on appropriate scaling. Otherwise, the classic BPINN formulation is prone to failure. The proposed adaptive weighting strategy avoids these issues altogether, performing much better in multi-scale inverse problems. 

We have demonstrated this in real-world applications from computational fluid mechanics (CFD) of incompressible flow in asymmetric 2D geometries. We showed the use of AW-HMC BPINNs for CFD inpainting and studies the impact of noise on the multi-potential energy. This highlighted the robustness of the present approach to noisy measurements, but also its ability to automatically adjust the $\lambda$ values to accurately estimate the noise levels themselves. In this sense, we were able to show enhanced uncertainty with increasing noise, without any prior on the noise level itself, and to capture distinct intrinsic task sensitivities to the noise. Overall, this offers an effective alternative to automatically address multi-fidelity problems with measurements resulting from unknown heteroscedastic noise distributions. 

Taken together, the present results render BPINNs a promising approach to scientific data assimilation. They now have the potential to effectively address multi-scale and multi-task inference problems, to couple UQ with physical priors, and to handle problems with sparse and noisy data. In all of these, the presented approach ensures efficient Pareto-front exploration, the ability to correctly scale multi-scale and stiff dynamics, and to derive unbiased uncertainty information from the data. Our approach involves only minimal assumptions on the noise distribution, the different problem scales, and the weights, and it is computationally efficient. This extends the application of BPINNs to more complex real-world problems that were previously not straightforwardly to address.  

Applications we expect to particularly benefit from these improvements include porous media research, systems biology, and the geosciences, where BPINNs now offer promising prospects for data-driven modeling. They could support and advance efforts for the extraction and prediction of morphological geometries \cite{phan_automatic_2021, Blunt_2021}, upscaling and coarse-graining of material properties \cite{alqahtani_machine_2020} and physical properties \cite{santos_computationally_2021} directly from sample images. However, capturing these features from imperfect images remains challenging and is usually subject to uncertainties, e.g., due to unavoidable imaging artifacts. This either requires the development of homogenization-based approaches \cite{HUME2021109910} to bridge scales and quantify these uncertainties \cite{perez_deviation_2022} or the use of data assimilation to compensate for the partial lack of knowledge in the images. The present BPINNs formulation with AW-HMC offers a potential solution.

\appendix
\normalsize

\section{Upper bound on the Inverse-Dirichlet weighting variance}
\label{sec:bound_Inv_Dir}

The Inverse-Dirichlet Adaptively Weighted HMC algorithm, developed in Sect. \ref{subsec:InvDir_AW_HMC}, guarantees that the gradients of the multi-potential energy terms have balanced distributions throughout the sampling, as shown by their joint variance below:
\begin{equation}
    \gamma^2 := \mathrm{Var}\{\lambda_k\nabla_\Theta \mathcal{L}_k \} \simeq \mathrm{min}_{t=0,...,K} (\mathrm{Var}\{\nabla_\Theta \mathcal{L}_t \}),\quad \forall k = 0,...,K.
\end{equation}
In this section, we use a general case to demonstrate that $\gamma^2$ is upper bound and controlled by a reliability criterion which depends on the prediction errors or PDE residuals, the dispersion of their mean variability with respect to $\Theta$ and the setting of the $\sigma_\bullet$ values. 

This first states the necessity to adequately set the $\sigma$ parameters to avoid biased and imbalanced conditions on task gradient distributions, since these parameters critically and arbitrarily affect the gradient distributions control. This also highlights that manual tuning of the $\sigma$ values may be an extremely sensitive task, difficult to achieve in practice. Therefore, in all the applications presented in this article, we chose  to set these parameters uniformly and instead rely on the $\lambda$ automatic adjustment to ensure, inter alia, the efficient exploration of the Pareto front.  It ensues that these standard deviation parameters imply a strong constraint on each gradient distribution — with respect to $\Theta$ — and so, impact each task uncertainty. 

For the sake of simplicity, we used two-task sampling with a data-fitting term from a field $u$ and a PDE constraint, denoted $\mathcal{F}$, so the data set is decomposed into $\mathcal{D} = \mathcal{D}^u \cup \mathcal{D}^\Omega$, following the notations introduced in Sect. \ref{subsec:MO_paradigm}. The multi-potential energy thus reduces to : 
\begin{equation}
    U(\Theta) = \frac{\lambda_0}{2\sigma_0^2}\left\|u_\Theta - u\right\|_{\mathcal{D}^u}^2
    + \frac{\lambda_1}{2\sigma_1^2}\left\|\mathcal{F}(u_\Theta)\right\|_{\mathcal{D}^\Omega}^2
    + \frac{1}{2\sigma_\Theta^2}\left\| \Theta\right\|^2 := \sum_{k=0}^{K+1} \lambda_k \mathcal{L}_k(\Theta)
\end{equation}
where we choose to keep the $\sigma$ notation for the demonstration and restrain $\Theta$ to the neural network parameters, even if the following holds in an inverse problem paradigm. As a reminder, the measurement data used for the training $\mathcal{D}^u$ can differ from the collocation points where we impose the PDE constraint $\mathcal{D}^\Omega$ and their respective numbers are denoted $N^u$ and $N^\Omega$. With the notations from Sect. \ref{subsec:MO_paradigm}, the gradients of the two-tasks potential energy write respectively: 
\begin{equation}
    \begin{aligned}
        \ds &\frac{\partial \mathcal{L}_0}{\partial \Theta_j}(\Theta) = \frac{1}{\sigma_0^2 N^u} \sum_{i=0}^{N^u} \bigg(u_\Theta(x_i) - u_i\bigg)\frac{\partial u_\Theta}{\partial \Theta_j}(x_i)\\
        &\frac{\partial \mathcal{L}_1}{\partial \Theta_j}(\Theta) = \frac{1}{\sigma_1^2 N^\Omega} \sum_{i=0}^{N^\Omega} \mathcal{F}\bigg(u_\Theta(x_i)\bigg)\frac{\partial\mathcal{F}(u_\Theta) }{\partial \Theta_j}(x_i)      
    \end{aligned}   
\end{equation}
for $\Theta\in \R^{p}$ and we can thus decompose the variances $\mathrm{Var}_\Theta\big[\nabla_\Theta \mathcal{L}_k\big],\, k=0,1$ with respect to these gradients. To do so, we first compute their mean with respect to $\Theta$ and get respectively  
\begin{equation}
    \ds \E_\Theta\big[\nabla_\Theta\mathcal{L}_0\big] = \frac{1}{N^{p}}\sum_{j=0}^{N^{p}}\frac{\partial\mathcal{L}_0}{\partial\Theta_j}(\Theta) = \frac{1}{\sigma_0^2 N^u} \sum_{i=0}^{N^u} \bigg(u_\Theta(x_i) - u_i\bigg) \E_\Theta\big[\nabla_\Theta u_\Theta\big](x_i)
    = \frac{1}{\sigma_0^2}  \E_{\mathcal{D}^u} \big[(u_\Theta - u) \E_\Theta \big[\nabla_\Theta u_\Theta\big]\big] 
\end{equation}
and 
\begin{equation}
     \E_\Theta\big[\nabla_\Theta\mathcal{L}_1\big] =\frac{1}{\sigma_1^2}  \E_{\mathcal{D}^\Omega} \big[\mathcal{F}(u_\Theta) \E_\Theta \big[\nabla_\Theta \mathcal{F}(u_\Theta) \big]\big]
\end{equation}
with the special configuration $\nabla_\Theta \mathcal{F}(u_\Theta) = \mathcal{F}(\nabla_\Theta u_\Theta)$ if $\mathcal{F}$ is linear. Finally, we can extend it to the variance computations, as follows:
\begin{equation}
\begin{aligned}
     \ds \mathrm{Var}_\Theta\big[\nabla_\Theta \mathcal{L}_0\big] &= \frac{1}{N^p} \sum_{j=0}^{N^{p}} \left( 
    \frac{\partial\mathcal{L}_0}{\partial\Theta_j} - \E_\Theta\big[\nabla_\Theta\mathcal{L}_0 \big]\right)^2\\
    &= \frac{1}{N^p(N^u\sigma_0^2)^2} \sum_{j=0}^{N^{p}} \left[ \sum_{i=0}^{N^{u}}\bigg(u_\Theta(x_i) - u_i\bigg) \bigg( \frac{\partial u_\Theta}{\partial\Theta_j}(x_i) -  \E_\Theta\big[\nabla_\Theta u_\Theta\big](x_i) \bigg)\right]^2\\
    &= \frac{1}{(N^u\sigma_0^2)^2}\sum_{i=0}^{N^{u}}\sum_{k=0}^{N^{u}}\bigg(u_\Theta(x_i) - u_i\bigg)\bigg(u_\Theta(x_k) - u_k\bigg) \mathrm{Cov}_\Theta\big[ \nabla_\Theta u_\Theta(x_i), \nabla_\Theta u_\Theta(x_k)  \big]\\
    &\leq \frac{1}{\sigma_0^4} \| u_\Theta -u\|^2_{\infty, \mathcal{D}^u} \mathrm{Cov}_\Theta\bigg[ \frac{1}{N^u}\sum_{i=0}^{N^u} \nabla_\Theta u_\Theta(x_i),\frac{1}{N^u}\sum_{k=0}^{N^u} \nabla_\Theta u_\Theta(x_k)  \bigg]\\
    &= \frac{1}{\sigma_0^4} \| u_\Theta -u\|^2_{\infty, \mathcal{D}^u} \mathrm{Var}_\Theta\big[ \E_{\mathcal{D}^u}\big[\nabla_\Theta u_\Theta \big] \big]
\end{aligned}
\end{equation}
that provides an upper bound for the gradient variance of the data-fitting term. We then obtain, in the same way, the PDE constraint bound as:
\begin{equation}
    \ds \mathrm{Var}_\Theta\big[\nabla_\Theta \mathcal{L}_1\big] \leq \frac{1}{\sigma_1^4} \| \mathcal{F}(u_\Theta)\|^2_{\infty, \mathcal{D}^\Omega} \mathrm{Var}_\Theta\big[ \E_{\mathcal{D}^\Omega}\big[\nabla_\Theta \mathcal{F}(u_\Theta) \big] \big].
\end{equation}
The notation $\|\cdot\|_{\infty,\mathcal{D}^\bullet}$ here refers to the discrete $\ell^\infty$ norm on the spatial domain composed of the $\mathcal{D}^\bullet$ training points, and $\E_{\mathcal{D}^\bullet}$ introduces the spatial mean on the corresponding data set. Hence, the gradient variances of the tasks are controlled by the crossed complex components $\mathrm{Var}_\Theta\E_{\mathcal{D}^\bullet}$ which can be interpreted as sensitivity terms evaluating the dispersion with respect to $\Theta$ of the gradient descent directions, averaged in space. Finally, since the $\sigma_\bullet$ values are uniformly set to one to avoid biased sampling, it means that the $\lambda$ values are computed in such a way the joint variance of the gradient distributions is bounded by: 
\begin{equation}
\label{upper_bound_gamma}
    \gamma^2\leq \mathrm{min} \left\{  \| u_\Theta -u\|^2_{\infty, \mathcal{D}^u} \mathrm{Var}_\Theta\big[ \E_{\mathcal{D}^u}\big[\nabla_\Theta u_\Theta \big] \big],\,  \| \mathcal{F}(u_\Theta)\|^2_{\infty, \mathcal{D}^\Omega} \mathrm{Var}_\Theta\big[ \E_{\mathcal{D}^\Omega}\big[\nabla_\Theta \mathcal{F}(u_\Theta) \big] \big]  \right\}
\end{equation}
which highlights the fact that the weights are adjusted with respect to the most likely task and thus improve the reliability in the uncertainty quantification. The present computations can straightforwardly be extended to more complex multi-potential energy terms for direct and inverse real-world problems, which concludes our analysis.
\section{2D Sobolev training benchmark}
\label{sec:2D_Sob_training}

\begin{figure}
    \centering
    \includegraphics[scale = 0.4]{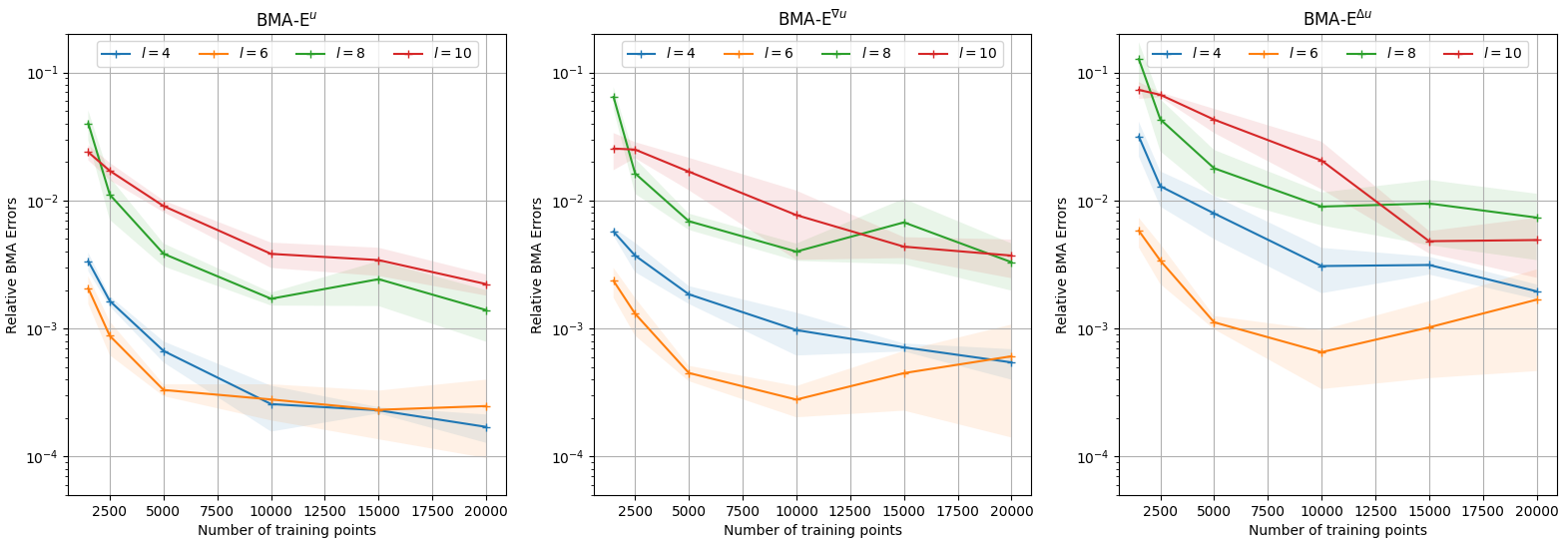}
    \caption{2D Sobolev training benchmark: comparison of the relative BMA errors, as defined in \eqref{BMA_rel_error}, plotted with respect to the number of training points for various shape complexities induced by the different values of $l$. The number of training points is increased until about $30\%$ of the whole data set is reached, for 20000 training points.}
    \label{fig:BMA_E_2D_Sob}
\end{figure}

\begin{figure}
    \centering
    \includegraphics[scale = 0.7]{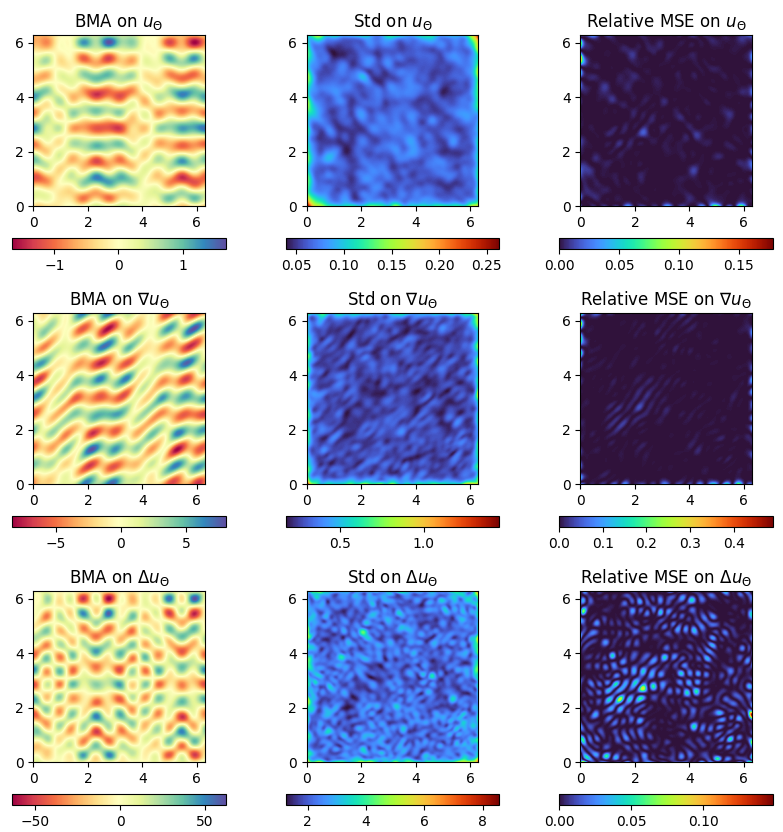}
    \caption{2D Sobolev training benchmark: BMA predictions, predicted standard deviation, and relative BMA errors, presented locally for each term of the multi-objective potential energy. We worked on a limited case $l=8$ with about $15\%$ of training points. The global relative BMA errors, averaged over the entire domain, scale around $1.69e-3$, $3.42e-3$, and $5.6e-3$ respectively.}
    \label{fig:2D_Sob_l8}
\end{figure}

We extend the Sobolev benchmark used in Sect.\ref{subsec:Sob_train_bench} to 2D-training with the gradient and laplacian operators, with a target functional in the form: 
\begin{equation}
    u(x,y) =  \sum_{i=1}^{N_{rep}} A_x^i \cos \left( 2 \pi L^{-1} l_x^i x + \phi_x^i\right)A_y^i \sin \left( 2 \pi L^{-1} l_y^i y + \phi_y^i\right)
\end{equation}
which enables us to deal with a wide range of shape complexities and sharp interfaces, in addition to the stiffness introduced by the higher-order derivatives. We set the domain size to $L = 2\pi $, the number of repetitions $N_{rep} = 5$, while the parameters $A_x$ and $A_y$ are independently and uniformly sampled from the interval $[-2,2]$, as are $\phi_x$ and $\phi_y$ from $[0,2\pi]$. In order to treat several shape complexities, we consider a range of parameter $l$ such that the local length scales $l_x$ and $l_y$ are randomly sampled from the set $\{ 1,2, ..., l\}$. The 2D spatial domain $[0,2\pi]^2$ is covered by a uniform grid with a resolution of $256\times 256$, along with randomly-selected training points. We then study both the impact of the functional complexity, by setting different values of $l$, and the number of training points on the Bayesian Model Average resulting from our AW-HMC methodology. 

The results on the entire benchmark setup, presented in Fig \ref{fig:BMA_E_2D_Sob}, show a convergence trend with an increasing number of training points, and this independently of the $l$ values, even though the relative BMA errors reach higher bounds with additional shape complexity. These relative BMA errors are computed according to equation \eqref{BMA_rel_error} and are average versions of different repetitions of Sobolev sampling, simultaneously running in parallel. In fact, in order to deal with the stochastic-induced process that may arise from the sampling variabilities themselves, we performed several realizations starting with distinct initializations of the neural network $\Theta^{t_0}$ and momentum $r^{t_0}$ parameters, which lead to different sampling realizations. We can potentially take into account these sampling variabilities to compute the standard deviation over these repetitions, as illustrated by the colored band in Fig \ref{fig:BMA_E_2D_Sob}. 

We also represent in Fig \ref{fig:2D_Sob_l8} for each term of the multi-potential functional, respectively, their BMA predictions, their uncertainties based on the predictive standard deviations throughout the sampling, and the relative BMA errors in the case $l=8$ and with $10000$ training points, randomly sampled over the whole domain. The results here show enhanced uncertainties near the boundary walls where the higher errors are located and highlight the ability of our methodology to capture complex shape fields of different orders of magnitude at the same time. We can also emphasize that such a 2D Sobolev training benchmark was previously unachievable with the classical BPINNs-HMC formulation.

\section{Failure of the usual methodologies on the Lokta-Voterra inverse problem}
\label{sec:LV_fails}

\begin{figure}
    \centering
    \includegraphics[scale = 0.7]{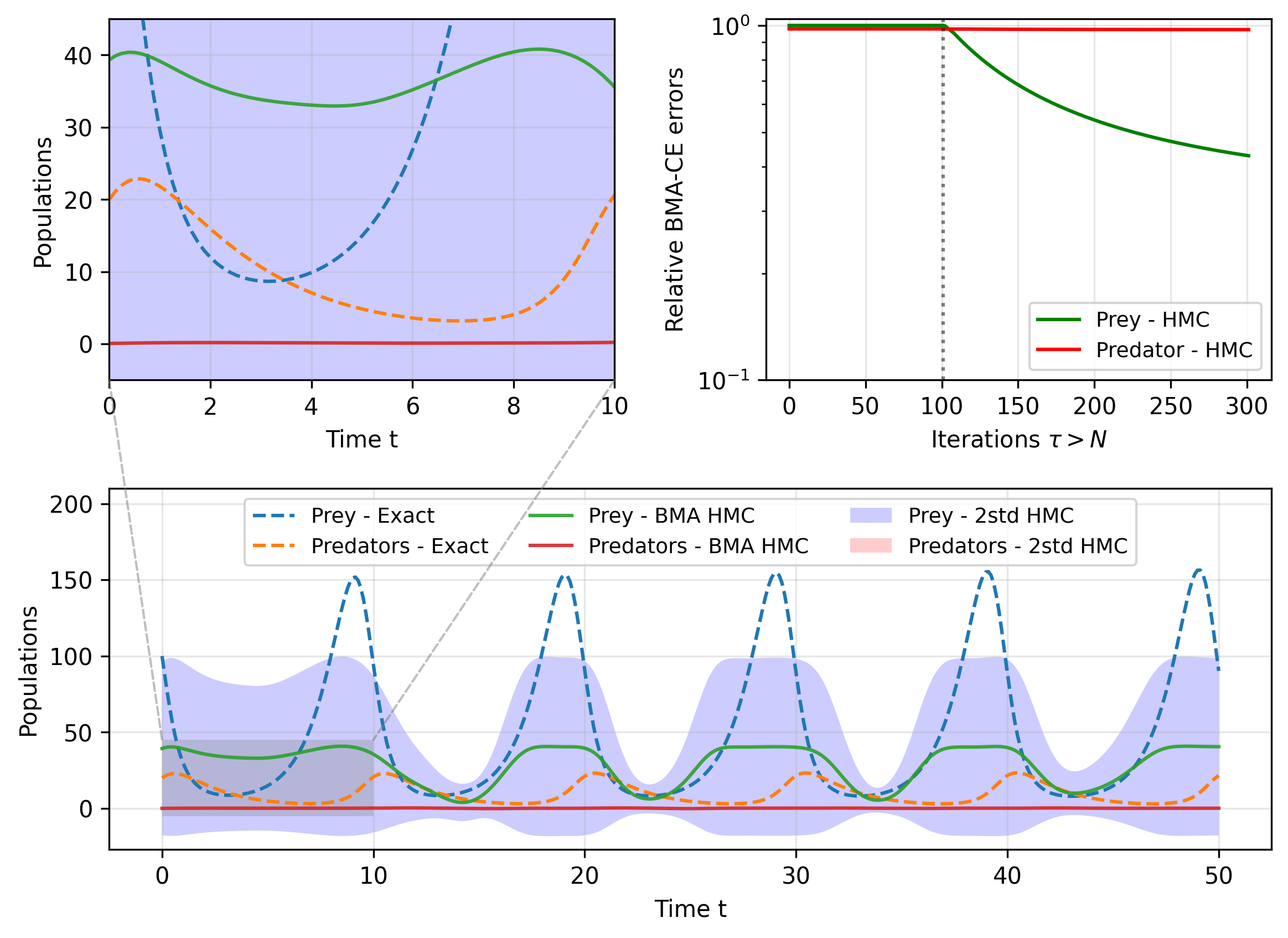}
    \caption{Failure mode of classical HMC, with uniform weighting, on the Lokta-Volterra multi-scale inverse problem defined in Sect. \ref{subsec:Lokta_Volterra}. BMA predictions for the two-species populations along the physical time with their uncertainties — bottom and top left figures. Relative BMA-CE errors throughout the sampling iterations illustrate lack of convergence of the method — top right.}
    \label{fig:LV_HMC_Fail}
\end{figure}

\begin{figure}
    \centering
    \includegraphics[scale = 0.7]{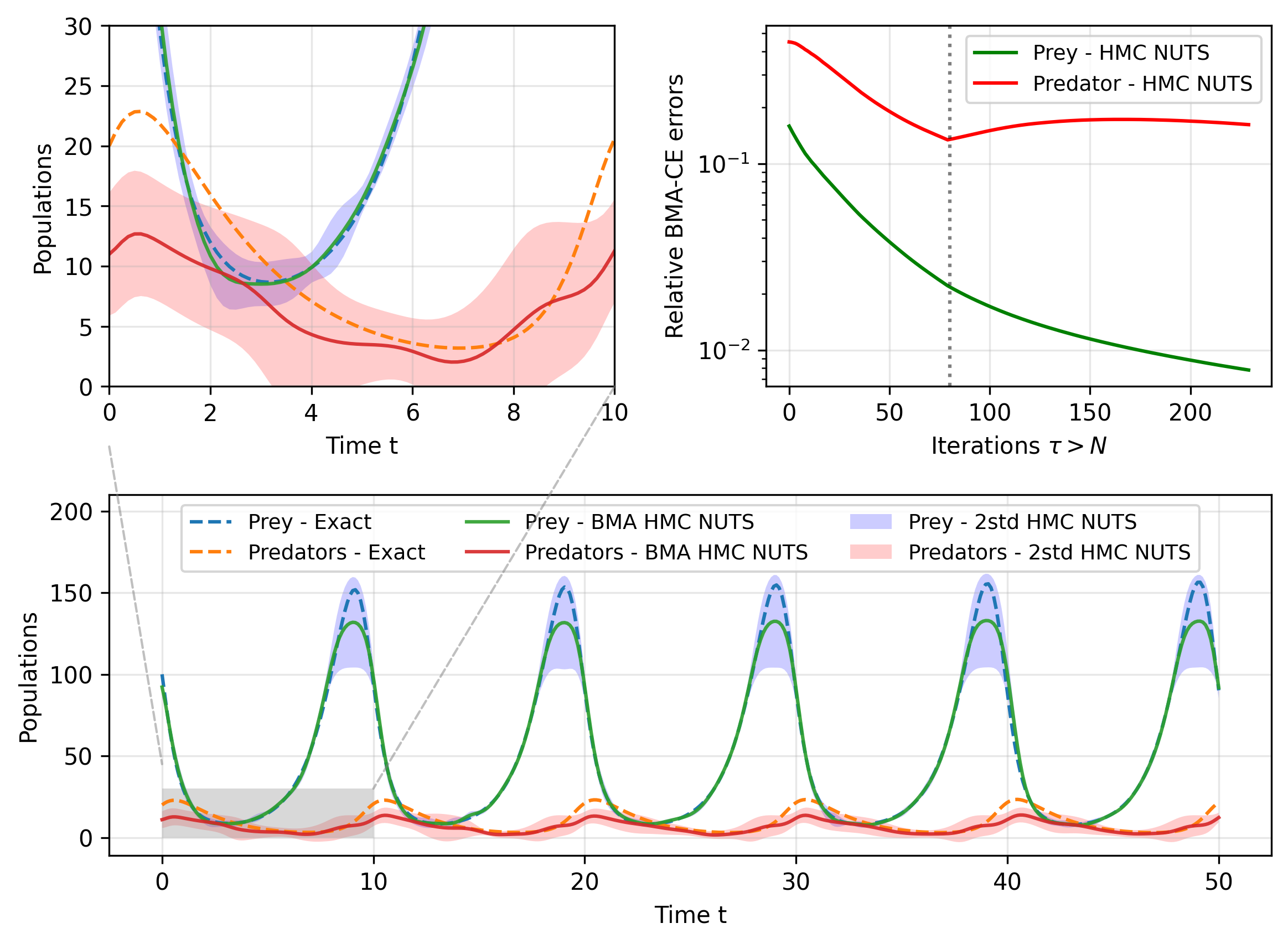}
    \caption{Failure mode of HMC with NUTS adaptation on the Lokta-Volterra multi-scale inverse problem defined in Sect. \ref{subsec:Lokta_Volterra}. BMA predictions for the two-species populations along the physical time with their uncertainties — bottom and top left figures. Relative BMA-CE errors throughout the sampling iterations showing an imbalance between the tasks and preferential adaptation of the prey population — top right figure.}
    \label{fig:LV_HMC_NUTS_Fail}
\end{figure}

\begin{figure}
    \centering
    \includegraphics[scale = 0.5]{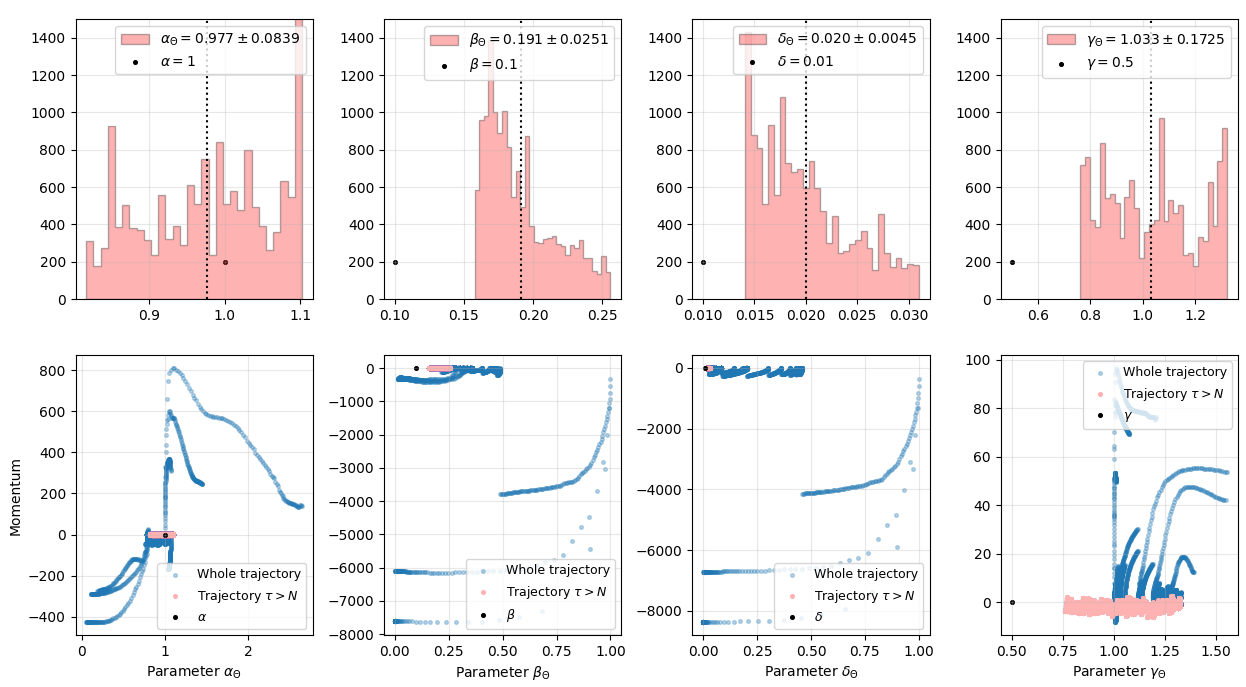}
    \caption{Failure mode of HMC with NUTS on the Lokta-Volterra multi-scale inference: histogram of the marginal posterior distributions for the inverse parameters (top) and phase diagrams of their trajectories throughout the sampling (bottom). The biased predictions from Fig \ref{fig:LV_HMC_NUTS_Fail} prevent proper inference of the inverse parameters, leading to random walk pathological behavior in the updated parameters.}
    \label{fig:LV_inv_params_Fail}
\end{figure}

We consider the Lokta-Volterra inverse problem, as introduced in Sect. \ref{subsec:Lokta_Volterra}, to investigate the impact of multi-scale dynamics on the usual methodologies, namely HMC with uniform weighting and NUTS. The sampling and leapfrog parameters are set accordingly to the AW-HMC test case, where $N$ refers to the burning and number of adaptive steps for the HMC and NUTS formulations, respectively. Therefore, we compare the different samplers assuming  that 1) their time complexity is the same and 2) we are imposing no informative priors on the inverse parameter scaling. In fact, the first condition states that different leapfrog parameters might improve the inference of these conventional methodologies. However, it implies a noticeable decrease in the leapfrog time step $\delta t$, thus slower exploration of the energy levels. Hence, these methods require either an increase in the integration time — by increasing $L$ — or using a large number of samples, to obtain suitable predictions. As a reminder, independently of this lack of efficiency in the posterior distribution sampling, poor choices on the weights of multi-potential energy can bias the sampler and deviate it from the Pareto front exploration. The second assumption is motivated by the willingness to address UQ on multi-scale dynamics without any prior knowledge of the separate scales. This arises from an assertion by Linka et al. \cite{linka_bayesian_2022} indicating that sensitivity to scaling disrupts the performance of the BPINNs-HMC. Finally, we also consider sequential training to provide an appropriate basis for comparison between the different methods. 

The results show a lack of convergence of the classical HMC with uniform weighting (in Fig \ref{fig:LV_HMC_Fail} — top right) and also, a strong imbalance between the tasks. The relative BMA-CE errors effectively characterize an extremely poor convergence of the predator population with respect to the prey population, which translates directly into an inefficient BMA prediction for the two-species populations (in Fig \ref{fig:LV_HMC_Fail} — bottom and top left). This failure mode is due essentially to the massive rejection of the samples (acceptance rate less than $1\%$) due to non-conservation of the Hamiltonian trajectories along the leapfrog steps. Hence, this confirms the lack of robustness of the BPINNs-HMC paradigm when facing instability issues due to multi-scale dynamics.

The NUTS alternative also struggles to converge on this multi-scale inverse problem and results in inadequate predictions, especially for the predator population. Here the reason is not the massive sample rejection but rather a prohibitive decrease in the time step, reaching $\delta t=\num{8.26e-5}$ and $\num{2.81e-5}$, respectively, at the end of the adaptive steps — nearly corresponding to a ten-fold drop in the time step, compared to AW-HMC. The relative BMA-CE errors (in Fig \ref{fig:LV_HMC_NUTS_Fail} — top right) reveal that this time step adaptation is suitable for the convergence of the prey population since it appears to be the most sensitive task. This sensitivity should be understood in the sense that small variations with respect to $\Theta$ on the potential energy induce the strongest constraint on the Hamiltonian energy conservation. 
However, the time-step adaptation is not satisfactory for the predator population and even leads to inefficient forgetting of the neural network throughout the sequential training. This translates into misleading predictions on the evolution of the population (see Fig \ref{fig:LV_HMC_NUTS_Fail} — bottom and top left) and unsuccessful inference of the inverse parameters (Fig \ref{fig:LV_inv_params_Fail}). The phase diagram of the inverse parameter trajectories demonstrates the difficulties of the NUTS sampler in adequately identifying the modes resulting from separate scales. Overall, the NUTS sampler suffers from a lack of convergence toward the Pareto front and a misleading inference of the inverse parameters, subject to weakly-informed priors, due to its inability to capture multi-scale behaviors.

\section{Characterization of the multi-potential energy in the CFD inverse problem}
\label{sec:inv_pb_potential}

The CFD inverse problem, defined in Sec \ref{subsec:Inv_pb}, involves the recovery of the latent pressure field $p_\Theta$ in addition to the flow regime parameter — given by the Reynolds number $Re_\Theta$ — based upon partial measurements of the velocity field. The training dataset $\mathcal{D}$ used for the AW-HMC sampling is first decomposed into 9559 measurements of randomly-sampled $\mathbf{u}$, which respectively defined the $\mathcal{D}^\mathbf{u}$ and $\mathcal{D}^\partial$ sets of interior and boundary points. The same collocation points define $\mathcal{D}^\Omega$, where we impose the PDE constraints and the diverge-free condition. The steep stenosis geometry considered in this problem generates sharp gradients at the wall interface. The latter need to be adequately captured to obtain consistency in the inference of the latent pressure and inverse parameter. Hence, we complemented the training with some partial measurements of the first-order derivatives of the velocity. This enables us to ensure that the convective terms, in the PDE constraints \eqref{NS_velocity}, are consistent with the velocity data and therefore infer the corresponding pressure field. The multi-potential energy is thus written as: 
\begin{equation}
\label{pot_inv_pb}
    \begin{aligned}
    U(\Theta) &= \frac{\lambda_0}{2\sigma_0^2}\left\|u_\Theta - u \right\|_{\mathcal{D}^{\mathbf{u}}}^2
    + \frac{\lambda_1}{2\sigma_1^2}\left\| v_\Theta - v \right\|_{\mathcal{D}^{\mathbf{u}}}^2
    + \frac{\lambda_2}{2\sigma_2^2}\left\|u_\Theta - u \right\|_{\mathcal{D}^{\partial}}^2
    + \frac{\lambda_3}{2\sigma_3^2}\left\| v_\Theta - v \right\|_{\mathcal{D}^{\partial}}^2 \\
    &+\frac{\lambda_4}{2\sigma_4^2}\left\|\partial_x u_\Theta - \partial_x u \right\|_{\mathcal{D}^{\mathbf{u}}}^2
    + \frac{\lambda_5}{2\sigma_5^2}\left\|\partial_x v_\Theta - \partial_x v \right\|_{\mathcal{D}^{\mathbf{u}}}^2
    + \frac{\lambda_6}{2\sigma_6^2}\left\|\partial_x u_\Theta - \partial_x u \right\|_{\mathcal{D}^{\partial}}^2 \\
    & + \frac{\lambda_7}{2\sigma_7^2}\left\|\partial_x v_\Theta - \partial_x v \right\|_{\mathcal{D}^{\partial}}^2
    + \frac{\lambda_8}{2\sigma_8^2}\left\| Re_\Theta^{-1}\Delta u_\Theta - (u_\Theta \partial_x u_\Theta + v_\Theta \partial_y u_\Theta) - \partial_x p_\Theta \right\|_{\mathcal{D}^{\Omega}}^2\\
    &+ \frac{\lambda_9}{2\sigma_9^2}\left\| Re_\Theta^{-1}\Delta v_\Theta - (u_\Theta \partial_x v_\Theta + v_\Theta \partial_y v_\Theta) - \partial_y p_\Theta \right\|_{\mathcal{D}^{\Omega}}^2\\ 
    &+ \frac{\lambda_{10}}{2\sigma_{10}^2}\left\|\nabla \cdot u_\Theta\right\|_{\mathcal{D}^{\Omega}}^2
     + \frac{\lambda_{11}}{2\sigma_{11}^2}\left\|\partial_y u_\Theta - \partial_y u \right\|_{\mathcal{D}^{\mathbf{u}}}^2
    + \frac{\lambda_{12}}{2\sigma_{12}^2}\left\|\partial_y u_\Theta - \partial_y u \right\|_{\mathcal{D}^{\partial}}^2 + \frac{1}{2\sigma_\Theta^2} \|\Theta\|^2_{R_{p+1}}
     \end{aligned}
\end{equation}
where the notation $\|\cdot\|$ refers to either the RMS norm on $\mathcal{D}^\bullet$ or the usual Euclidean norm on $\R^{p+1}$.

\end{document}